\documentclass[aps,prl,twocolumn,reprint,
			   groupedaddress,superscriptaddress,
			   amsfonts,amssymb,amsmath,
			   citeautoscript,
			   a4paper,footinbib
			   ]{revtex4-1}

\usepackage{tikz}
\usepackage{ulem}
\usepackage{microtype} %For better kerning and symbol-stretching

\usepackage{txfonts}  %Times Roman fonts
\usepackage{txfontsb} %Addition for txfonts, including old style numerals and greek
\usepackage{bm} %Bold math symbols with \bm{} (Greek and other symbols)

\usepackage{xcolor}
\usepackage{graphicx}

\usepackage{physics}
\usepackage{hyperref}
\hypersetup{
    colorlinks,
    linkcolor={blue!75!black!80!yellow},
    citecolor={blue!75!black!80!yellow},
    urlcolor={blue!75!black!80!yellow}
}

\usepackage[centering,hmargin=1.85cm,tmargin=31mm,bmargin=37mm]{geometry}
  \usepackage{wrapfig, framed}
\color{black!88!white}

\newcommand{\nv}{\hat{\mathbf n}}
\newcommand{\rv}{\mathbf r}

\newcommand{\rvb}{\rv_{\scriptscriptstyle \partial V_{\rm res}}}
\newcommand{\Ev}{{\mathbf E}}
\newcommand{\Pv}{{\mathbf P}}

\newcommand{\Iv}{{\bf{\sf\displaystyle I}}}

\newcommand{\nablav}{{\bm{\nabla}}}

\newcommand{\greenOP}{{\bf{\sf\displaystyle G}}}
\newcommand\reduline{\bgroup\markoverwith
{\textcolor{red}{\rule[-0.5ex]{2pt}{0.4pt}}}\ULon}

\definecolor{reviseC}{rgb}{0.23   0.23 0.7   }

\begin{document}
%-----TITLE-----
\title{{\color{black!88!white}
Shape deformation of nanoresonator: a quasinormal-mode perturbation theory
}}

%-----AUTHORS AND AFFILIATIONS-----
\author{Wei~Yan}
\email{wyanzju@gmail.com}
\affiliation{Key Laboratory of 3D Micro/Nano Fabrication and Characterization of Zhejiang Province, School of Engineering, Westlake University, 18 Shilongshan Road, Hangzhou 310024, Zhejiang Province, China}
\affiliation{Institute of Advanced Technology, Westlake Institute for Advanced Study, 18 Shilongshan Road, Hangzhou 310024, Zhejiang Province, China}

\author{Philippe~Lalanne}
\email{philippe.lalanne@institutoptique.fr}
\affiliation{Laboratoire Photonique, Num{\'e}rique et Nanosciences (LP2N), IOGS-Univ. Bordeaux-CNRS, 33400 Talence cedex, France}

\author{Min~Qiu}
\email{qiumin@westlake.edu.cn}
\affiliation{Key Laboratory of 3D Micro/Nano Fabrication and Characterization of Zhejiang Province, School of Engineering, Westlake University, 18 Shilongshan Road, Hangzhou 310024, Zhejiang Province, China}
\affiliation{Institute of Advanced Technology, Westlake Institute for Advanced Study, 18 Shilongshan Road, Hangzhou 310024, Zhejiang Province, China}

\date{\today}
%

%--------------------
%----- ABSTRACT -----
%--------------------
\begin{abstract}\color{black!88!white}

When material parameters are fixed, optical responses of nanoresonators are dictated by their shapes and dimensions. Therefore, both designing nanoresonators and understanding their underlying physics would benefit from a theory that predicts the evolutions of resonance modes of open systems---the so-called quasinormal modes (QNMs)---as the nanoresonator shape changes. QNM perturbation theories (PTs) are one ideal choice. However, existing theories developed for material changes are unable to provide accurate perturbation corrections for shape deformations. By introducing a novel extrapolation technique, we develop a rigorous QNM PT that faithfully represents the electromagnetic fields in perturbed domain. Numerical tests performed on the eigenfrequencies, eigenmodes and optical responses of deformed nanoresonators evidence the predictive force of the present PT, even for \textit{large deformations}. This opens new avenues for inverse design, as we exemplify by designing super-cavity modes and exceptional points with remarkable ease and physical insight.

\end{abstract}

%---------------------
%----- MAKETITLE -----
%---------------------
\maketitle

\color{black!88!white}
%-----------------------
%----- Introduction-----
%-----------------------

Plasmonc and Mie nanoresonators that confine light in tiny volumes play an essential role in nanophotonics~\cite{Novotny:2012}. Their modelling requires full-wave simulations~\cite{Lalanne:2018}, and accordingly their design is computationally expensive, even with advanced inverse design algorithms~\cite{Jensen:2011,Sean:2018} that smartly explore parameter space for repeated wave-excitation instances. Complementary approaches, with a better balance between physics and numerics, are desirable.

Here lies the worth of cavity perturbation theory (PT), a well-known principle permeating various branches of physics, which predicts resonances of new (perturbed) problems from resonances of an initial (unperturbed) one~\cite{Kato:1976}. For tiny perturbations, accurate predictions of frequency shifts of individual modes are delivered with single-mode first-order PTs. For large perturbations, if a complete set of unperturbed modes is known, exact solutions can be, in principle, obtained with the modal superposition method. Initial contributions on cavity PTs rely on Hermitian formalisms (normal modes), along with a crucial technique, hereafter called as the local-field correction (LFC), which increases the accuracy of unknown (perturbed) modal fields by incorporating quasi-static depolarization fields in perturbation region~\cite{Harrington:1961,Johnson:2002}. However, these initial Hermitian formalisms are strictly valid only for closed systems, hardly legitimate for high-$Q$ dielectric resonators and largely inconsistent for low-$Q$ nanoresonators, see~\cite{Yang:2015} and Sec. S2 of~\cite{SeeSupplementalMaterial}.
They have to be replaced by non-Hermitian formalisms based on resonance modes of open systems, the so-called quasinormal modes (QNMs)~\footnote{{For a review of the impact of non-Hermiticity on first-order cavity PT and other related physical phenomena, e.g. Purcell effects, please refer to~\cite{Lalanne:2018}}}.

Owing to issues on the basis completeness, non-Hermitian cavity PTs have been mainly devoted to permittivity changes inside resonator in seminal works~\cite{Leung:1994b,Lee:1999b} and more recent ones~\cite{Muljarov:2016b}, or to minute permittivity changes outside the resonator~\cite{Weiss:2016}.
The important case of shape deformations---of great practical interest for design---involving both inward and outward perturbations has received comparatively minimal attention. Only tiny deformations have been considered so far in the restricted case of single-mode PTs~\cite{Leung:1990,Yang:2015}.

In this letter, capitalizing on these earlier works, we address this shortcoming and propose a rigorous non-Hermitian PT framework for shape deformations. The framework is established on an advanced modal basis that combines a restricted set of dominant QNMs with additional numerical modes~\cite{ Vial:2014,Yan:2018}. This physically preserves the insight of QNM expansions and mathematically guaranties the completeness of the modal expansion in the interior and exterior of the resonator. The framework additionally benefits from a completely novel extrapolation technique that provides a faithful representation of the perturbed modes in the perturbed region and naturally implements the LFC at arbitrary perturbation order. The extrapolation technique enables the derivation of exact formulas for both first- and high-order perturbation corrections and plays an essential role in the reported superior performance. As shown by numerical tests, large deformations, with volume changes of typically 30\%-50\% and high metal-dielectric permittivity contrasts, can be accurately handled with a modest number of modes. As outlined in the last part, our results open new perspectives for inverse design in nanophotonics, a topic wherein brute-force computation is insufficient and supplemental theoretical insights are needed~\cite{Miller:2014,Wu:2020}.

%%%%%%%%%%%%%%%%%%%%%%%%%%%%%%%%%%%%%%%%%%%%%%%%%%%%%%%%%%%%%%%%%%%%%%%%%%%%%%%%%%%%%%%%%%%%%%%%%%%%%%%%%%%%%%%%%%%%%%%%%%%%%%%%%%%%%%%%%%%%%%%%%%%%%%%%%%%%%%%%%%%%%%%%%%%%%%%%%%%%%%%%%%%%%%%

\begin{figure}[!t]
\centering
\includegraphics[width=8.7cm]{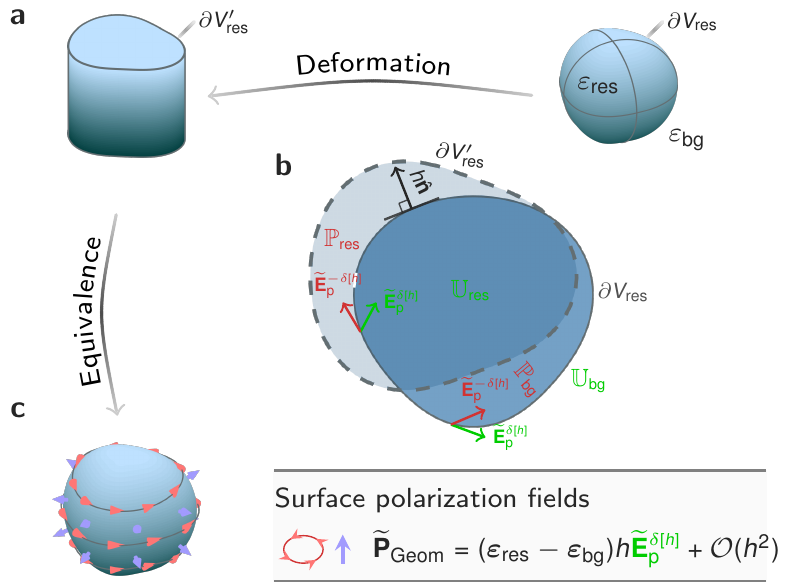}
\caption{{\bf Overview of the PT framework and notations.} {\bf a.} A body (nanoresonator) is deformed with boundary changing from $\partial V_{\rm res}$ to $\partial V_{\rm res}'$. {\bf b.} Geometrical deformation is parameterized by $h(\rvb)\nv$.
%Perturbed domains are labelled by $\bm {\mathbb{P}}_{\rm res, bg}$, while unperturbed domains are labelled by ${\mathbb{U}}_{\rm res, bg}$.
%The perturbed domains, where material parameters change after the deformation, are labelled by $\bm {\mathbb{P}}_{\rm res, bg}$ with the subscripts implying that they belong to the perturbed resonator (res) or background (bg), respectively. Similarly, the unperturbed domains are labelled by $\bm
{\bf c.} The perturbed body is modeled as the unperturbed body dressed (augmented) by a surface-polarization distribution, $\widetilde{\mathbf P}_{\rm Geom}$ given by Eq. \eqref{eq:PGEOM}. In {\bf b-c}, $\widetilde\Ev^{\pm\scriptscriptstyle \delta[h]}\equiv\widetilde\Ev(\rv_{\scriptscriptstyle \partial V_{\rm res}}\pm \delta[h])$ with $\delta[h]\equiv 0^-$ for $h>0$ and otherwise $\delta[h]\equiv 0^+$.
}
\label{Fig:1}
\end{figure}

Figure 1 shows our notations. A body (nanoresonator) is deformed into a perturbed one with its boundary changing from
$\partial V_{\rm res}$ to $\partial V{'}_{\rm res}$. The deformation is parameterized by $h(\rvb)\nv$ measuring perpendicular shift from $\partial V_{\rm res}$ to $\partial V_{\rm res}'$, where $\nv$ denotes unit outward normal vector of $\partial V_{\rm res}$ and $\rvb$ denotes coordinates on $\partial V_{\rm res}$. The permittivity tensors of the nanoresonator and background are denoted by $\bm\varepsilon_{\rm res}$ and $\bm\varepsilon_{\rm bg}$, respectively. Outward ($h>0$) and inward ($h<0$) deformations result in material changes $\Delta\bm\varepsilon\equiv\bm\varepsilon_{\rm res}-\bm\varepsilon_{\rm bg}$ and  $-\Delta\bm\varepsilon$, respectively, in perturbed domains denoted by ${\mathbb{P}}_{\rm res}$ and ${\mathbb{P}}_{\rm bg}$, respectively.
The remaining unperturbed domain is denoted by ${\mathbb{U}}_{\rm res, bg}$ belonging either to the resonator (res) or background (bg).

%%%%%%%%%%%%%%%%%%%%%%%%%%%%%%%%%%%%%%%%%%%%%%%%%%%%%%%%%%%%%%%%%%%%%%%%%%%%%%%%%%%%%%%%%%%%%%%%%%%%%%%%%%%%%%%%%%%%%%%%%%%%%%%%%%%%%%%%%%%%%%%%%%%%%%%%%%%%%
%%%%%%%%%%%%%%%%%%%%%%%%%%%%%%%%%%%%%%%%%%%%%%%%%%%%%%%%%%%%%%%%%%%%%%%%%%%%%%%%%%%%%%%%%%%%%%%%%%%%%%%%%%%%%%%%%%%%%%%%%%%%%%%%%%%%%%%%%%%%%%%%%%%%%%%%%%%%%%%%%%%%%%%%%%%%%%%%%%%%%%%%%%%%%%%

{\bf Field Extrapolation in Perturbed Domains}: We start from the Lippman–Schwinger integral equation expressing the electric fields of the perturbed modes,
$\widetilde \Ev_{\rm p}$, with the Green's tensor of the unperturbed system, $\greenOP_{\rm u}$:
$\widetilde \Ev_{\rm p}(\rv)=\omega^2 \int \greenOP_{\rm u}(\rv,\rv';\omega) f(\rv')\Delta\bm\varepsilon (\omega)\widetilde\Ev_{\rm p}(\rv') d^3\rv'$~\cite{SeeSupplementalMaterial}
, where $f(\rv)$ is a filling function with values of $1$ and $\textnormal{-}1$ for $\rv\in {\mathbb{P}}_{\rm res}$ and ${\mathbb{P}}_{\rm bg}$, respectively, and $0$ elsewhere.

To expand the perturbed modes $\widetilde\Ev_{{\rm p}}$ with a complete basis for both inward and outward deformations, we consider a set of unperturbed modes $\widetilde\Ev_{{\rm u}}$ composed of a subset of dominant QNMs and additional numerical modes~\cite{Vial:2014}. Highly accurate reconstructions in this modal basis have been recently obtained for complex problems, involving noncompact shapes (e.g. resonator dimers) and nonuniform environments (e.g. metallic substrates)~\cite{Yan:2018}, as well as gratings with their many inevitable branch cuts in the complex-frequency plane~\cite{Gras:2019}. However, for shape deformations, directly expanding $\widetilde\Ev_{{\rm p}}$ into the $\widetilde\Ev_{{\rm u}}$ basis would lead to nonuniform convergence owing to field discontinuity across $\partial V_{\rm es}$, see~\cite{Leung:1990,Johnson:2002}. To bypass this issue in a systematic way, we here develop a novel extrapolation technique that allows us to consider large deformations. First, disregarding ${\mathbb{P}}_{\rm res, bg}$ domains, we perform the modal expansion in the ${\mathbb{U}}_{\rm res, bg}$ domains only, $\widetilde \Ev_{\rm p}(\rv)=\sum_ n \alpha_n\widetilde \Ev_{{\rm u};n}(\rv)$, $\alpha_n$ being the expansion coefficient. Then, we take a { key} step and extrapolate $\widetilde \Ev_{\rm p}$ in $\mathbb{P}_{\rm res,bg}$ from fields in $\mathbb{U}_{\rm res,bg }$ with a Taylor expansion of $\widetilde \Ev_{\rm p}$ about $\partial V_{\rm res}$:
$
\widetilde\Ev_{\rm p}(\rv)=\sum_{j=0}^\infty (l^j/j!) {\overrightarrow{\partial}}_{\nv}^{j}\widetilde\Ev_{\rm p}(\rv_{\scriptscriptstyle \partial V_{\rm res}}+\delta[h]\nv)
$. Here $\delta[h]\equiv 0^{-}$ for $h>0$ and otherwise $\delta[h]\equiv 0^{+}$; $\rv = l\nv+\rv_{\scriptscriptstyle \partial V_{\rm res}}$ with $l\in[0,\,h]$;
${\overrightarrow{\partial}}_{\nv}^j f(\rv_{\scriptscriptstyle \partial V_{\rm res}})\equiv(\nv\cdot \nablav)^jf(\rv_{\scriptscriptstyle \partial V_{\rm res}})$. The Taylor expansion is justified because the materials in ${\mathbb{P}_{\rm res, bg}}$ are the same as in ${\mathbb{U}_{\rm res, bg}}$ and electric fields in uniform domains (without permittivity discontinuities) are analytic.

%%%%%%%%%%%%%%%%%%%%%%%%%%%%%%%%%%%%%%%%%%%%%%%%%%%%%%%%%%%%%%%%%%%%%%%%%%%%%%%%%%%%%%%%%%%%%%%%%%%%%%%%%%%%%%%%%%%%%%%%%%%%%%%%%%%%%%%%%%%%%%%%%%%%%%%%%%%%%%%%%%%%%%%%%%%%%%%%%%%%%%%%%%%%%%%

The volume-integral Lippman–Schwinger equation is then reformulated as a surface-integral equation over $\partial V_{\rm res}$~\cite{SeeSupplementalMaterial}:
\begin{subequations}
\begin{align}
\widetilde\Ev_{\rm p}(\rv){=} {\oint_{\partial V_{\rm res}}}\greenOP_{\rm u}(\rv,\rvb{-}{\delta[h]}\nv;\omega)\widetilde\Pv_{\rm Geom}(\rv_{\scriptscriptstyle \partial V_{\rm res}}) d^2\rvb,
\label{eq:L-S2}
\end{align}
with surface polarization $\Pv_{\rm Geom}$ given by
\begin{align}
\widetilde\Pv_{\rm Geom}(\rv_{\scriptscriptstyle \partial V_{\rm res}})=\sum_{k=0}^{\infty}\sum_{j=0}^{\infty} {\overleftarrow{\partial}}_{\nv}^{\,\,k}
\Delta\bm\varepsilon\,
{\rm c}_{\scriptscriptstyle jk} {\overrightarrow{\partial}}_{\nv}^{j}\widetilde \Ev_{\rm p} (\rv_{\scriptscriptstyle \partial V_{\rm res}}{+}{\delta[h]\nv}).
\label{eq:PGEOM}
\end{align}
Here ${\rm c}_{\scriptscriptstyle jk}=\frac{h^{k+j+1}}{k!j!}\left(
\frac{1}{k+j+1}+\kappa_{\rm m}\frac{2h}{k+j+2}+
\kappa_{\rm g}\frac{h^2}{k+j+3}\right)$ with
$\kappa_{\rm m,g}$ denoting the mean and Gaussian curvatures of $\partial V_{\rm res}$, respectively; $f(\rv_{\scriptscriptstyle \partial V_{\rm res}}){\overleftarrow{\partial}}_{\nv}^{\,\,k}\equiv(\nv\cdot \nablav)^kf(\rv_{\scriptscriptstyle \partial V_{\rm res}})$.
\label{eq:nnew}
\end{subequations}

\begin{figure}[!t]
\centering
\includegraphics[width=8.7cm]{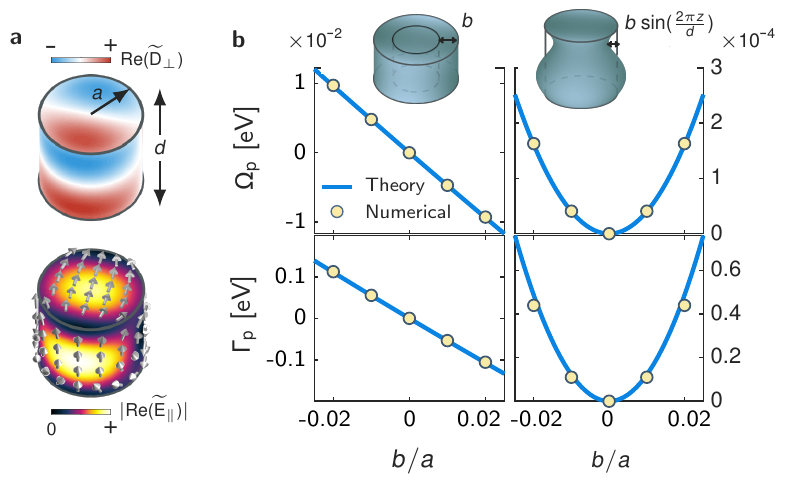}
\caption{{\bf Validation of the perturbation theory} for a silicon rod (radius $a = 200\, \rm nm$, height $d = 400 \,\rm nm$ and permittivity $\varepsilon_{\rm si}=12.96$) in air.
%a cross-sectional radius change $h=b\cos(m\phi)$, where $m$ is an integer and $\phi$ measures the radial angle.
{\bf a.} Field distribution of a (1,1,1) Mie's mode on the rod surface. The QNM frequency is $\widetilde \omega_{\rm u}=0.74-0.034{\rm i}\,\rm eV$. (Top) Real part of perpendicular normalized displacement field; (bottom) amplitude of real part of (vectorial) parallel normalized electric fields. The arrows specify the field direction.
{\bf b.} Modal eigenfrequency shifts, $\Delta\widetilde\omega\equiv\Delta\widetilde\Omega-{\rm i}\Delta\widetilde\Gamma/2$, as the nanorod radius varies either uniformly (left, $h=b$) or sinusoidally  [right,  $h=b\sin(2\pi z/d)$]. The (left) linear and (right) quadratic dependencies of $\Delta\widetilde\omega$ on deformation parameter $b$ are accurately predicted using the first-order and second-order PTs, respectively.
}
\label{Fig:OR}
\end{figure}

Equations \eqref{eq:nnew} are the cornerstone of our lately developed PT. They define a new integral formulation for electric fields of perturbed modes, which shall allow us to conveniently derive PT formulae to arbitrary orders.

{\bf Perturbation Theory}: Injection the modal expansions of $\widetilde\Ev_{\rm p}$ and
$\greenOP_{\rm u}$~\cite{Lalanne:2018} in Eqs.~\eqref{eq:nnew}, we obtain a linear eigenvalue equation for perturbed {modes} $\left\{\widetilde\omega_{\rm p},\ket{\bm\alpha}\equiv[\alpha_1;\alpha_2;\cdots]\right\}$ (Sec. S5 of ~\cite{SeeSupplementalMaterial}):
\begin{subequations}
\begin{align}
{\bf\mathcal{H}}_0 \ket{\bm\alpha}=\widetilde\omega_{\rm p}\left[\Iv+{\bf\mathcal{H}}_{\rm p}\right] \ket{\bm\alpha}.
\label{eq:PT}
\end{align}
Here ${\bf\mathcal{H}}_0$ is a diagonal matrix with diagonal elements being frequencies of unperturbed modes;
$\Iv$ denotes the identity matrix. ${\bf\mathcal{H}}_{\rm p}$ accounts for the perturbation contribution, for which
we make the first-order approximation
\begin{align}
{
\mathcal{H}_{{\rm p};nm}\simeq\mel{\left(\widetilde\Ev_{{\rm u};n}^{\rm bg}\right)^*}{ h\Delta\bm\varepsilon(\widetilde\omega_{{\rm u};m})}{\widetilde\Ev_{{\rm u};m}^{\rm res}}_{\partial V_{\rm res}}},
\label{eq:H1st}
\end{align}
where $\mel{\left(\widetilde\Ev _{{\rm u};n}^{\rm bg}\right)^*}{h\Delta\bm\varepsilon(\widetilde\omega_{{\rm u};m})}{\widetilde\Ev _{{\rm u};m}^{\rm res}}_{\partial V_{\rm res}}{\equiv}\oint_{\partial V_{\rm res}} \widetilde\Ev _{{\rm u};n}(\rv_{\scriptscriptstyle \partial V_{\rm res}}{+}{0^+}\nv)\cdot
{h(\rv_{\scriptscriptstyle \partial V_{\rm res}})\Delta\bm\varepsilon(\widetilde\omega_{{\rm u};m})}
\cdot \widetilde\Ev _{{\rm u};m} (\rv_{\scriptscriptstyle \partial V_{\rm res}}{+}{0^-}\nv)d^2\rvb$, and $\widetilde\Ev _{{\rm u};n}^{\rm bg}$ and
$\widetilde\Ev _{{\rm u};n}^{\rm res}$ denote $\widetilde\Ev _{{\rm u};n}$ at the outer and inner sides of $\partial V_{\rm res}$, respectively.
\label{eq:MainResult}
\end{subequations}

Equations \eqref{eq:MainResult} constitute our first important result, a rigorous first-order PT for deformation problems of open systems.
%They merge two novelties---the Taylor expansion of perturbed QNM fields and
%an advanced formalism combining dominant QNMs and additional numerical modes.
The essential difference with earlier works~\cite{Muljarov:2016b,Weiss:2016} is the interplay of the inner and outer fields at the resonator boundary in the perturbation matrix ${\bf\mathcal{H}}_{\rm p}$. The interplay guaranties that for vanishing $h$, $\widetilde \Ev_{{\rm p}; n}$ uniformly converges towards $\widetilde \Ev_{{\rm u}; n}$ for all $n$---whereas earlier formalisms do so only nonuniformly, see Sec. S2 of \cite{SeeSupplementalMaterial}---, thereby allowing us to obtain accurate predictions for large deformations with a small number of retained QNMs. Accordingly, it is unnecessary to include numerical modes practically (at least for the first-order PT corrections).
For small deformations, the first-order single-mode frequency shift, $\Delta\widetilde\omega_{n}\equiv\widetilde\omega_{{\rm p}; n}-\widetilde\omega_{{\rm u}; n}$, is given by
\begin{align}
\Delta\widetilde\omega_{n}\simeq  -\widetilde\omega_{{\rm u};n}\mel{\left(\widetilde\Ev_{{\rm u};n}^{\rm bg}\right)^*}{ h\Delta\bm\varepsilon(\widetilde\omega_{{\rm u};n})}{\widetilde\Ev_{{\rm u};n}^{\rm res}}_{\partial V_{\rm res}},
\label{eq:1st-fs}
\end{align}
which is consistent with earlier works on normal-mode PTs~\cite{Johnson:2002}---in the limit of ${\rm Im}(\widetilde\omega_{{\rm u};n})\to 0$ and ${\rm Im}(\widetilde\Ev_{{\rm u};n})\to 0$---
and QNM PTs using the LFC for tiny deformations~\cite{Leung:1990,Yang:2015}.

{\bf Validation:} We consider a silicon rod in air that supports Mie's resonances indexed by $(q,n,l)$---the azimuthal, radial and longitudinal numbers. Figure \ref{Fig:OR}a shows the field distribution of a $(1,1,1)$ mode, which is selected for the following study. In this initial study aiming at the validation of the first-order PT of Eq.~\eqref{eq:MainResult}, the rod is only slightly deformed by a uniform radial change $h=b$. The left panel in Fig.~\ref{Fig:OR}b compares the values of $\Delta\widetilde\omega$ predicted from Eq.~\eqref{eq:1st-fs} with exact numerical data obtained with the QNMEig solver of the freeware MAN (Modal Analysis of Nanoresonators)~\cite{Yan:2018,Anu:2013} implemented with COMSOL Multiphysics. The quantitative agreement, along with similar observations in Figs. S2-S3~\cite{SeeSupplementalMaterial}, evidences the soundness of Eq.~\eqref{eq:1st-fs} in the limit of vanishing perturbations.

The first-order PT can be generalized to high-order ones by retaining high-order terms in ${\bf\mathcal{H}}_{\rm p}$ when solving the eigenvalue problem of Eq. \eqref{eq:PT}. As an example to validate our theory at high order, we consider another radial deformation $h=b\sin(2\pi z/d)$, where $z$ denotes the longitudinal coordinate. In this case, the first-order correction vanishes since $h$ is odd with respect to $z$, and the second-order correction is dominant.
As evidenced in the right panel of Fig.~\ref{Fig:OR}b and also in Figs. S4 and S5~\cite{SeeSupplementalMaterial}, the second-order PT accurately predicts the quadratic frequency shift $\Delta\widetilde\omega$ of the $(1,1,1)$ mode, with a residual error due to modal truncation \footnote{{The computation of $\Delta\widetilde\omega$ with the second-order PT includes modes with $|\widetilde\omega_{\rm u}|<10\,[\rm eV]$.}}. {Compared to first-order results, the accuracy improvement is obvious. However, the second-order PT requires a much larger number of modes to reach the accuracy. Balancing between accuracy and effectiveness suggests us to use the first-order PT for the following studies.

\begin{figure}[!t]
\centering
\includegraphics[width=8.7cm]{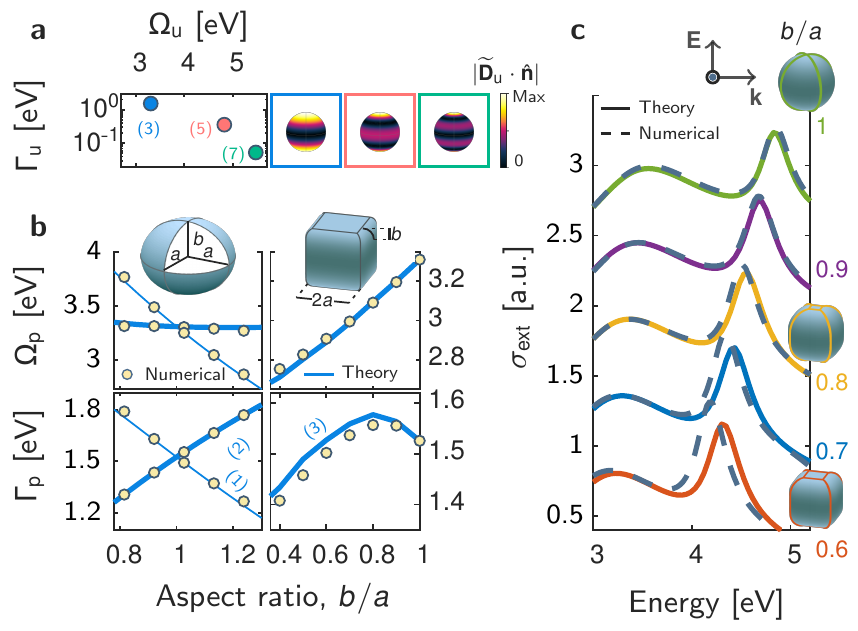}
\caption{{\bf Test for large deformations} of a silver sphere (50-nm radius) in air. Two deformations, into spheroids or cuboids, are considered. Silver is modelled by a Drude permittivity $\varepsilon_{\rm Ag}=1-\omega_p^2/(\omega^2+i\omega\gamma)$ with $\hbar\omega_p=9$ eV and $\hbar\gamma=0.021$ eV.
{\bf a.} (Left) Eigenfrequencies $\widetilde\omega_{\rm u}\equiv\Omega_{\rm u}-i\Gamma_{\rm u}/2$ of dipole (blue) and quadrupole (red) QNMs. Degeneracy factors are given in parenthesis. (Right) Amplitudes of perpendicular normalized electric displacement fields for dipole, quadrupole and hexapole QNMs with azimuthal order $m=0$.
{\bf b.} Eigenfrequencies $\widetilde\omega_{\rm p}\equiv\Omega_{\rm p}-i\Gamma_{\rm p}/2$ of perturbed dipole QNMs for spheroids and cuboids as aspect ratios $b/a$ vary (for small deformations, $b/a \simeq 1$). {\bf c.} Extinction-cross-section spectra of cuboids for several values of $b/a$.
$2\times 15$ QNMs are used in {\bf b-c}; additional static QNMs at zero frequency are taken into account in $\bf c$.
}
\label{Fig:2}
\end{figure}
\label{eq:New}

\begin{figure*}[!htb]
\centering
\includegraphics[width=12.4cm]{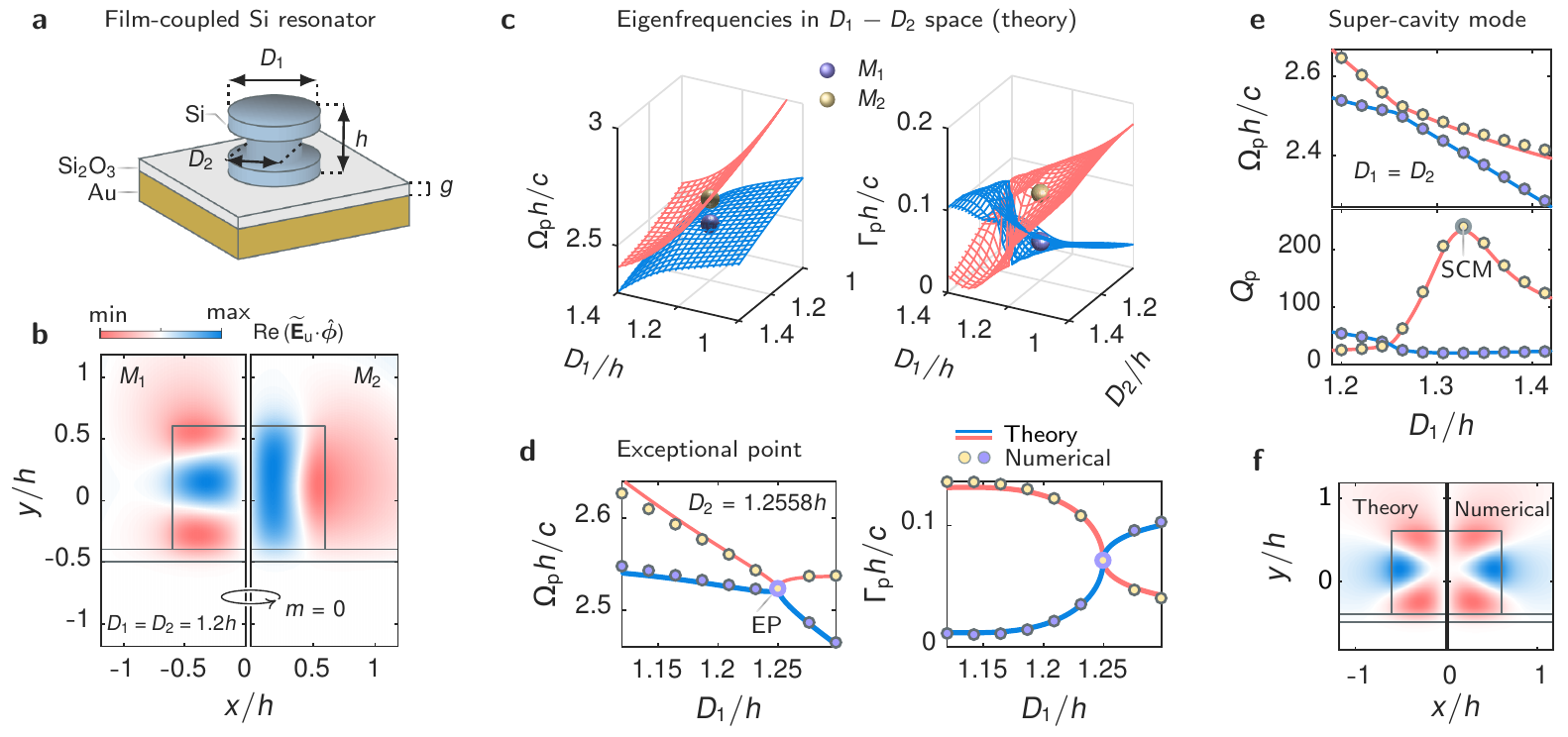}
\caption{{\bf Application of the PT for designing exceptional point (EP) and high-$Q$ super-cavity mode (SCM).} {\bf a.} A Si dumbbell-shaped nanoresonator sits on an Au substrate coated with a 50-nm-thick $\text{SiO}_2$ layer. The dumbbell consists of three equal-high cylinders (total height $h=500\,\rm nm$) with diameters $D_1$, $D_2$ and $D_1$ (from top to bottom).
The Au permittivity is approximated by the Lorentz-Drude model
$\varepsilon_{\rm Au}(\omega)=\varepsilon_{\infty}-\omega_{p,1}^2/(\omega^2+i\gamma_1\omega)-\omega_{p,2}^2/(\omega^2-\omega_0^2+i\gamma_2\omega)$ with
$\varepsilon_{\infty}=6$, $\hbar\omega_{p,1}= 8.67\, \rm eV$, $\hbar\gamma_{1}= 0.1\, \rm eV$,  $\hbar\omega_{p,2}=3.65\, \rm eV$, $\hbar\gamma_{2}= 2.15\, \rm eV$, $\hbar\omega_0=7.38\, \rm eV$.
{\bf b.} Field distributions of azimuthal-component electric fields, ${\rm Re} \left(\widetilde E_{\rm u;\phi }\right)$, of two modes $M_1$ and $M_2$ with azimuthal order $m=0$ obtained for $D_1=D_2=1.2 h$.
{\bf c.} Eigenfrequencies calculated with Eqs. \eqref{eq:MainResult} for perturbed QNMs resulting from $M_1-M_2$ coupling as the parameter space ($D_1-D_2$) is spanned.
{\bf d.} Eigenfrequencies for $D_2=1.2558 h$. The EP is labelled.
{\bf e.} Resonance frequencies and quality factors for $D_1=D_2$. The SCM with $Q_{\rm p}\simeq 240$ is labelled in the lower panel.
{\bf f.} SCM-field distribution ${\rm Re}\left (\widetilde E_{\rm p;\phi }\right)$.
}
\label{Fig:3}
\end{figure*}

{\bf Reconstruction of scattered fields:} The use of the frequency shift formula of Eqs.~\eqref{eq:MainResult} or ~\eqref{eq:1st-fs} for designing nanoresonators with tailored resonance wavelengths or quality factors will be discussed later.
In inverse design, the possibility of predicting the nanoresonator response with unperturbed modes is equally useful~\cite{Jensen:2011,Sean:2018,Miller:2013}. Thus we consider a perturbed nanoresonator driven by an incident field $\Ev_{\rm in}$ and denote by $\Ev_{\rm sca}$ the scattered field.
By taking into account volume polarization $\Delta\bm\varepsilon\Ev_{\rm in}$ due to
the incident field, Eqs. \eqref{eq:nnew} is generalized: $\Ev_{\rm sca}(\rv)=\int_{V_{\rm res}} \greenOP_{\rm u}(\rv,\rv';\omega)\Delta\bm\varepsilon(\omega)\Ev_{\rm in}(\rv') d^3\rv'+ \oint_{\partial V_{\rm res}} \greenOP_{\rm u}(\rv,\rv_{\scriptscriptstyle \partial V_{\rm res}}'-\delta\nv;\omega) \Pv_{\rm Geom}(\rv_{\scriptscriptstyle \partial V_{\rm res}}') d^2\rv_{\scriptscriptstyle \partial V_{\rm res}}'$ (see Sec. S8 of ~\cite{SeeSupplementalMaterial}) where $\Pv_{\rm Geom}$ is expressed with Eq.~\eqref{eq:PGEOM} using the total field $\Ev_{\rm in}+\Ev_{\rm sca}$.
By expanding $\Ev_{\rm sca}$ with unperturbed modes, $\Ev_{\rm sca}=\sum_n \beta_n \widetilde\Ev_{{\rm u};n}$, and performing the modal expansion for $\greenOP_{\rm u}$, we obtain a linear equation for $\ket{\bm\beta}\equiv[\beta_1;\beta_2;\cdots]$:
\begin{align}
{\bf\mathcal{H}}_0 \ket{\bm\beta}=\omega\left[\Iv+{\bf\mathcal{H}}_{\rm p}\right] \ket{\bm\beta}+\omega \left[\ket{\bm B}+ \ket{\bm S}\right],
\label{eq:beta}
\end{align}
where the source terms $\ket{\bm B}\equiv[B_1;B_2;\cdots]$ and $\ket{\bm S}\equiv[S_1;S_2;\cdots]$ with
$B_n=\int_{V_{\rm res}} \widetilde\Ev_{{\rm u};n}(\rv) \Delta\bm\varepsilon(\widetilde\omega_n) \Ev_{\rm in}(\rv) d^3\rv $
and
$
S_n=\mel{\left(\widetilde\Ev_{{\rm u};n}^{\rm bg}\right)^*}{
h \bm\Delta\varepsilon(\widetilde\omega_n)
}{\Ev_{\rm in}}_{\scriptscriptstyle \partial V_{\rm res}}
$. Note that, for $\Ev_{\rm in}=0$ ($\ket{\bm S}=\ket{\bm B}=0$), Eq. \eqref{eq:beta} reduces to the eigenvalue equation for perturbed modes; when the perturbation vanishes (${\bf\mathcal{H}}_{\rm p}=0$ and $\ket{\bm S}=0$), $\ket{\bm \beta}$'s become modal excitation coefficients of the unperturbed nanoresonator. Equation \eqref{eq:beta} allows us to reconstruct optical responses of perturbed nanoresonators and constitutes the {\it second} main result of this letter.

%%%%%%%%%%%%%%%%%%%%%%%%%%%%%%%%%%%%%%%%%%%%%%%%%%%%%%%%%%%%%%%%%%%%%%%%%%%%%%%%%%%%%%%%%%%%%%%%%%%%%%%%%%%%%%%%%%%%%%%%%%%%%%%%%%%%%%%%%%%%%%%%%%%%%%%%%%%%%%%%%%%%%%%%%%%%%%%%%%%%%%%%%%
%%%%%%%%%%%%%%%%%%%%%%%%%%%%%%%%%%%%%%%%%%%%%%%%%%%%%%%%%%%%%%%%%%%%%%%%%%%%%%%%%%%%%%%%%%%%%%%%%%%%%%%%%%%%%%%%%%%%%%%%%%%%%%%%%%%%%%%%%%%%%%%%%%%%%%%%%%%%%%%%%%%%%%%%%%%%%%%%%%%%%%%%%%

{\bf Application:} When performing inverse design of a photonic device, one explores a large parameter space to optimize several electromagnetic observables with typically gradient based algorithms through repeated simulations of Maxwell’s equations. The developed PT offers new opportunities for geometrical optimization. First, QNM expansions make the physics transparent, thereby helping the interpretation of
optimized results. Second, the computational cost of problems involving broad bandwidth, multi-frequency bands or multi-illumination instances can be dramatically reduced~\cite{Lalanne:2018}, benefiting from the analyticity of objective functions. Third, since nanoresonator responses are generally driven by a few QNMs, the optimization problem in large parameter space becomes more tractable~\cite{Sean:2018, Wu:2020}, and the expensive task of computing gradients, with either finite schemes or the adjoint method, is simplified due to small-dimensional matrix problems of Eqs.~\eqref{eq:MainResult} or \eqref{eq:beta}. How far these equations may allow us to accurately explore parameter space---before being obliged to locally restabilize the optimization by computing again a few dominant QNMs---decisively impact the effectiveness of the present PT.

}

%%%%%%%%%%%%%%%%%%%%%%%%%%%%%%%%%%%%%%%%%%%%%%%%%%%%%%%%%%%%%%%%%%%%%%%%%%%%%%%%%%%%%%%%%%%%%%%%%%%%%%%%%%%%%%%%%%%%%%%%%%%%%%%%%%%%%%%%%%%%%%%%%%%%%%%%%%%%%%%%%%%%%%%%%%%%%%%%%%%%%%%%%%
%%%%%%%%%%%%%%%%%%%%%%%%%%%%%%%%%%%%%%%%%%%%%%%%%%%%%%%%%%%%%%%%%%%%%%%%%%%%%%%%%%%%%%%%%%%%%%%%%%%%%%%%%%%%%%%%%%%%%%%%%%%%%%%%%%%%%%%%%%%%%%%%%%%%%%%%%%%%%%%%%%%%%%%%%%%%%%%%%%%%%%%%%%

To quantify the exploration capability of Eqs. \eqref{eq:MainResult} that take into account deformation-induced couplings between different modes, we consider large spheroidal and cuboidal deformations of a silver sphere. The results are summarized in Fig. \ref{Fig:2}. For solving Eq.~\eqref{eq:MainResult}, 15 modes---whose frequencies and modal profiles are shown in Fig. \ref{Fig:2}a---plus their 15 complex conjugated counterparts $\left\{-\widetilde\omega_{{\rm u}; n}^*,\widetilde\Ev_{{\rm u};n}^*\right\}$ are considered, thereby giving a $30 \times 30$ ${\bf\mathcal{H}}_{\rm p}$ matrix. Figure \ref{Fig:2}b compares the theoretical predictions of the fundamental-dipole-QNM frequencies of the deformed geometries with the numerical data. Note that, for spheroids, the original dipole triplet is split into a doublet and a singlet. An overall quantitative agreement, up to deformations with 30\%-50\% volume changes, is achieved. This level of accuracy for large deformations, using a few QNMs, is largely unattainable with available cavity QNM \cite{Muljarov:2010, Weiss:2016} or normal-mode PTs ~\cite{Johnson:2002} (see the comparisons in Fig. S1~\cite{SeeSupplementalMaterial}). In Fig.~\ref{Fig:2}c, we additionally compare the theoretical predictions of Eq.~\eqref{eq:beta} for the extinction-cross-section spectra of cuboids with exact numerical results obtained with the boundary element method, showing again quantitative agreement. More numerical evidences are shown in Figs. S6-9~\cite{SeeSupplementalMaterial}.

We further exemplify the potential of the present PT for inverse design by designing super-cavity modes (SCMs) and exceptional points (EPs) (a general workflow of employing the PT for inverse design is detailed in Sec. S9 of ~\cite{SeeSupplementalMaterial}).
SCMs, the analogues of bound states in the continuum for finite-size structures, offer high $Q$’s owing to destructive radiation interferences~\cite{Rybin:2017}, while EPs with two or more coalescing states have implications for lasing~\cite{Wong:2014} and sensing~\cite{Chen:2017} applications. Hereafter, we consider a complex geometry, a Si dumbbell-shaped resonator deposited on an Au substrate coated with a thin ${\rm Si}{\rm O}_2$ film. Since SCMs and EPs can be constructed with a bi-mode coupled system by carefully tuning modal coupling constants, we restrict the parameter space to two diameters, $D_1$ and $D_2$ (see Fig. \ref{Fig:3}a).
The design begins with a guessed geometry, $D_1=D_2=600\,\rm nm$ and $h=500\,\rm nm$ (height), for which we compute the QNMs with the solver QNMEig~\cite{Yan:2018}. We further select two QNMs, denoted by $M_1$ and $M_2$---that are frequency-protected from others, i.e., $|\widetilde\omega_{{\rm u};n}-\widetilde\omega_{{\rm u};M_i}|\gg|{\bf \mathcal{H}}_{{\rm p}; nM_i}|$ for $n\notin\left\{M_1,M_2\right\}$---, thereby defining an isolated bi-mode coupled system. Figure \ref{Fig:3}b shows the modal profiles of $M_{1,2}$ with azimuthal order $m=0$. Now, the calculation of the perturbed QNMs with Eqs.~\eqref{eq:MainResult} amounts to solve a simple $2\times 2$ eigenmatrix. As shown with Fig. \ref{Fig:3}c, we can directly and straightforwardly explore the entire parameter space with Eqs. \eqref{eq:MainResult}, without requiring any further time-consuming full-wave computations of the QNMs.

An EP is obtained for $D_2=1.2558\,h$ when the two eigenvalues coalesce as $D_1$ is varied, see details in Fig. 4d. On the other hand, the design of SCMs, revealed by their high-$Q$ values, does not necessitate a shape optimization as precise as that for EPs. For instance, constraining $D_1=D_2$, we observe in Fig.~\ref{Fig:3}e that $Q_{\rm p}$ of one mode is significantly increased for $D_1=650\,\rm nm$, identifying a SCM with a 20-fold $Q$ enhancement (see Fig. \ref{Fig:3}f for the mode profiles). Again, no need for further iterative full-wave computations; the SCM is directly found by exploring the parameter space with Eq.~\eqref{eq:MainResult}. Additionally note the quantitative agreement between the theoretical predictions and full-wave numerical data; and, accordingly, further optimization iterations are thus unnecessary.

%%%%%%%%%%%%%%%%%%%%%%%%%%%%%%%%%%%%%%%%%%%%%%%%%%%%%%%%%%%%%%%%%%%%%%%%%%%%%%%%%%%%%%%%%%%%%%%%%%%%%%%%%%%%%%%%%%%%%%%%%%%%%%%%%%%%%%%%%%%%%%%%%%%%%%%%%%%%%%%%%%%%%%%%%%%%%%%%%%%%%%%%%%%%%
%%%%%%%%%%%%%%%%%%%%%%%%%%%%%%%%%%%%%%%%%%%%%%%%%%%%%%%%%%%%%%%%%%%%%%%%%%%%%%%%%%%%%%%%%%%%%%%%%%%%%%%%%%%%%%%%%%%%%%%%%%%%%%%%%%%%%%%%%%%%%%%%%%%%%%%%%%%%%%%%%%%%%%%%%%%%%%%%%%%%%%%%%%%%%

{\bf Conclusions:} The present PT establishes a general and rigorous framework for predicting the optical responses of largely deformed resonators from the sole knowledge of the initial unperturbed modes. There is, in principle, no restriction on the resonator geometry and constitutive materials. It offers unprecedented numerical efficiency and physical transparency, making it a good tool for nanoresonator design.

{\bf Acknowledgements}---This project was supported by the National Key Research and Development Program of China (2017YFA0205700), the National Natural Science Foundation of China (61927820).

\bibliographystyle{apsrev4-2}
\bibliography{references}

\end{document}

%-----TITLE-----
\title{\color{black!88!white} SUPPLEMENTARY INFORMATION\vskip .5em
Shape deformation of nanoresonator: a quasinormal-mode perturbation theory}

%-----AUTHORS AND AFFILIATIONS-----
\author{Wei~Yan}
\email{wyanzju@gmail.com}
\affiliation{Key Laboratory of 3D Micro/Nano Fabrication and Characterization of Zhejiang Province, School of Engineering, Westlake University, 18 Shilongshan Road, Hangzhou 310024, Zhejiang Province, China}
\affiliation{Institute of Advanced Technology, Westlake Institute for Advanced Study, 18 Shilongshan Road, Hangzhou 310024, Zhejiang Province, China}

\author{Philippe~Lalanne}
\email{philippe.lalanne@institutoptique.fr}
\affiliation{Laboratoire Photonique, Num{\'e}rique et Nanosciences (LP2N), IOGS-Univ. Bordeaux-CNRS, 33400 Talence cedex, France}

\author{Min~Qiu}
\email{qiumin@westlake.edu.cn}
\affiliation{Key Laboratory of 3D Micro/Nano Fabrication and Characterization of Zhejiang Province, School of Engineering, Westlake University, 18 Shilongshan Road, Hangzhou 310024, Zhejiang Province, China}
\affiliation{Institute of Advanced Technology, Westlake Institute for Advanced Study, 18 Shilongshan Road, Hangzhou 310024, Zhejiang Province, China}

%---------------------
%----- MAKETITLE -----
%---------------------
\maketitle

\color{black!88!white}

%-----------------------------
%----- TABLE OF CONTENTS -----
%-----------------------------

\noindent{\small\textbf{\textsf{CONTENTS}}}\\
\twocolumngrid
\begin{spacing}{1}
{
\begingroup %The table of contents is typeset in bold for some reason - we want to avoid this, so for the moment we relax bfseries to do nothing.
\let\bfseries\relax
\let\tocdepth\relax

\tableofcontents
\endgroup
}
\end{spacing}

%-----------------------
%----- MAIN MATTER -----
%-----------------------

\onecolumngrid

\section{Notations and Definitions}
\label{Sec:Nota}

\begin{itemize}

  \item An unperturbed system consists of a resonator and a background medium.

  \begin{itemize}
  \item $\bm \varepsilon_{\rm res}$ and $\bm \varepsilon_{\rm bg}$ denote the permittivity tensors of the resonator and background, respectively.

  \item $\widetilde\Ev_{{\rm u};n}$ and $\widetilde\omega_{{\rm u};n}$ ($n=1,2,\cdots$) denote modal electric fields and frequencies, respectively.

  \item $\partial V_{\rm res}$ denotes the boundary separating the resonator and background, whose unit normal vector (pointing outward the resonator) is denoted by $\nv$.

  \item  $\rvb$ denotes the coordinates on $\partial V_{\rm res}$.
        \end{itemize}

   \item Shifting its boundary $\partial V_{\rm res}$ to $\partial V'_{\rm res}$, the unperturbed nanoresonator is deformed into a perturbed one.

    \begin{itemize}
    \item $h(\rvb)\nv$ parameterizes perpendicular shift from $\partial V_{\rm res}$ to $\partial V'_{\rm res}$.
    \item $\widetilde\Ev_{{\rm p};n}$ and $\widetilde\omega_{{\rm p};n}$ ($n=1,2,\cdots$) denote perturbed modal electric fields and frequencies, respectively.
    \item The domain between by $\partial V_{\rm res}$ and $\partial V'_{\rm res}$ is perturbed domain---where material parameters change due to the deformation---, and is labelled by $\mathbb{P}_{\rm res, bg}$ with their subscripts implying that they belong either to the resonator (res) or background (bg). $\mathbb{U}_{\rm res, bg}$ denote unperturbed domains.
    \end{itemize}

%\item The QNM perturbation theory (PT) is established by extrapolating $\widetilde\Ev_{{\rm p};n}$ in $\mathbb{P}_{\rm res, bg}$ domains from $\widetilde\Ev_{{\rm p};n}$ in $\mathbb{U}_{\rm res, bg}$ domains. The latter is expanded with the QNMs of the unperturbed system, $\widetilde\Ev_{{\rm u};n}$.
%\begin{siderules}
%\begin{enumerate}
% \item In $\mathbb{P}_{\rm res, bg}$ domains, $\rv\equiv \rvb +l \nv$ with $l\in[0,\;h]$.
%  \item $\rvb^{\sigma[h]}\equiv\rvb+\sigma[h]\nv$, where $\sigma[h]\equiv 0^-$ (negative infinitesimal) for $h>0$  and otherwise
%$\sigma[h]\equiv 0^+$ (negative infinitesimal). Similarly, $\rvb^{-\sigma[h]}\equiv\rvb-\sigma[h]\nv$.
%Following the above settings, $\rvb^{\rm bg}=\rvb^{0+}$ and $\rvb^{\rm res}=\rvb^{0-}$, thus, specify the background and resonator sides of $\partial V_{\rm res}$, respectively.
%\item   $\mathbf f^{\pm \sigma[h]}\equiv \mathbf f(\rvb^{\pm \sigma[h]}) $
%     and $\mathbf f^{\rm bg, res}\equiv \mathbf f(\rvb^{\rm bg, res}) $.
%    \item ${\overrightarrow{\partial}}_{\nv}^{j} \mathbf f(\rvb) \equiv (\nv\cdot \nabla)^j \mathbf f(\rvb)$;
%    $\mathbf f(\rvb){\overleftarrow{\partial}}_{\nv}^{j}  \equiv (\nv\cdot \nabla)^j \mathbf f(\rvb)$, where $j$ is a positive integer.
%\end{enumerate}
%\end{siderules}

\end{itemize}

\section{Comparison with Other First-order Perturbation Theories}

This section aims at explaining the difference between the present formalism and at evidencing its enhanced capability compared to earlier works.
We here only discuss formalisms that have addressed multi-mode first-order PTs. We omit single-mode PTs~\cite{ Harrington:1961,Klein:1993,Yang:2015,Weiss:2016,Cognee:2019}, which are restricted to the analysis of tiny perturbations. However, in the relation with the following discussion, note important contributions on single-mode PTs, like for instance, comprehensive studies on the non-Hermitian PT predictions for low-$Q$ nanoresonators~\cite{Yang:2015} and for high-$Q$ photonic-crystal resonators~\cite{Cognee:2019}.

\begin{enumerate}
\item \emph{Resonant-state expansion} PT with QNMs has been initially introduced in electromagnetism in the seminal works of Refs.~\cite{Leung:1994b,Lee:1999b}, and then further developed by Muljarov and colleagues, see Ref.~\cite{Muljarov:2010,Muljarov:2016b} and the series of related papers by the same group, under the name of resonant-state expansion (RSE). According to these works~\cite{Muljarov:2010,Muljarov:2016b}, electric fields in perturbed domains are directly expanded with unperturbed QNMs, therein leading to a first-order-perturbation Hamiltonian $\mathcal{H}_{\rm p} $ (using the notation of the present work) expression
    \begin{align}
      \mathcal{H}_{{\rm p}; nm} \simeq
       \mel{\left(\widetilde\Ev_{{\rm u};n}^{-\sigma[h]}\right)^*}{
       h \Delta\bm\varepsilon(\widetilde\omega_{{\rm u};m})}{\widetilde\Ev_{{\rm u};m}^{-\sigma[h]}}_{\scriptscriptstyle \partial V_{\rm res}},
      \label{eq:H1storder_OTH1}
     \end{align}
where $\widetilde\Ev_{{\rm u};n}^{-\sigma[h]}\equiv \widetilde\Ev_{{\rm u};n}(\rvb-\delta[h]\nv)$ with $\delta[h]=0^-$ for $h>0$ and otherwise $\delta[h]=0^+$.

In contrast with the present theory, Eq.~(2b) in the main text, Eq.~\eqref{eq:H1storder_OTH1} involves fields on a single ({\it perturbed}) side of the resonator boundary $\partial V_{\rm res}$, not on both sides. This is the result of expanding perturbed fields directly with unperturbed QNMs. Moreover, since QNMs only form a complete basis inside the resonator~\cite{Lalanne:2018}, the RSE approach is, by essence, restricted to perturbation changes inside the resonator, and in principle cannot handle the important case of shape deformations.

To analytically evidence that this formalism fails to predict the correct first-order frequency shift $\Delta\widetilde\omega_{n}\equiv\widetilde\omega_{{\rm p};n}-\widetilde\omega_{{\rm u};n}$ for deformation problems,
we simply focus on the single-mode first-order PT expression [as directly derived from Eq.~\eqref{eq:H1storder_OTH1}]
\begin{align}
     \Delta\widetilde\omega_{n}= -\widetilde\omega_{{\rm u};n}\mel{\left(\widetilde\Ev_{{\rm u};n}^{-\sigma[h]}\right)^*}{ h\Delta\bm\varepsilon(\widetilde\omega_{{\rm u};n})}{\widetilde\Ev_{{\rm u};n}^{-\sigma[h]}}_{\scriptscriptstyle \partial V_{\rm res}}.
     \label{eq:1stCPT}
    \end{align}
    and further consider a concrete case: A nanosphere that is deformed from an initial radius $r_0$ to $r_1$, i.e., $h=r_1-r_0$. Depending on the sign of $h$, $\Delta\widetilde\omega_{n}$ takes the different expressions
    \begin{align}
           \Delta\widetilde\omega_{n}=\left\{
                  \begin{array}{ll}
                  -h\mel{\left(\widetilde\Ev_{{\rm u};n}^{\rm{bg}}\right)^*}{\Delta\bm\varepsilon(\widetilde\omega_{{\rm u};n})}{\widetilde\Ev_{{\rm u};n}^{\rm {bg}}}_{\partial\Omega_u}\;\;\;\;h>0\\
                  -h\mel{\left(\widetilde\Ev_{{\rm u};n}^{\rm{res}}\right)^*}{\Delta\bm\varepsilon(\widetilde\omega_{{\rm u};n})}{\widetilde\Ev_{{\rm u};n}^{\rm {res}}}_{\partial\Omega_u}\;\;\;\;h<0
                 \end{array}
         \right
         .,
    \end{align}
    where $\widetilde\Ev_{{\rm u};n}^{\rm{bg}}$ and $\widetilde\Ev_{{\rm u};n}^{\rm{res}}$ denote $\widetilde\Ev_{{\rm u};n}$ at the outer (background) and inner (resonator) sides of $\partial V_{\rm res}$.
    As long as $\widetilde\Ev_{{\rm u};n}^{\rm {bg}}$ and
    $\widetilde\Ev_{{\rm u};n}^{\rm {res}}$ are different, the above equations indicate that the first-order derivative of $\Delta\widetilde\omega_{n}$ with respect to $h$ is discontinuous at $h=0$. Thus $\Delta\widetilde\omega_{n}$ is not analytic and, accordingly, Eq. \eqref{eq:1stCPT} is incorrect for analyzing shape deformations.

The incorrectness/inadequacy has a drastic impact on the numerical performance. To evidence this, we consider the numerical example of Fig. 3 in the main text, a silver nanosphere deformed into spheroids. The results of the comparison between the RSE PT, the present formalism and the exact numerical data are shown in Fig.~ \ref{Fig:1stDiffC}. We see that the predictions of the RSE PT are largely inaccurate, even when the geometrical deformation parameter $b/a$ is close to the non-perturbed value $b/a=1$. In order to achieve accurate predictions with the RSE PT, one needs to consider inward deformations (towards the interior of the resonator) for the sake of the QNM completeness and to retain a very large number of QNMs in the expansion to accommodate for the discontinuities introduced at the boundary.

\begin{figure}[!htp]
\centering
\includegraphics[width=10cm]{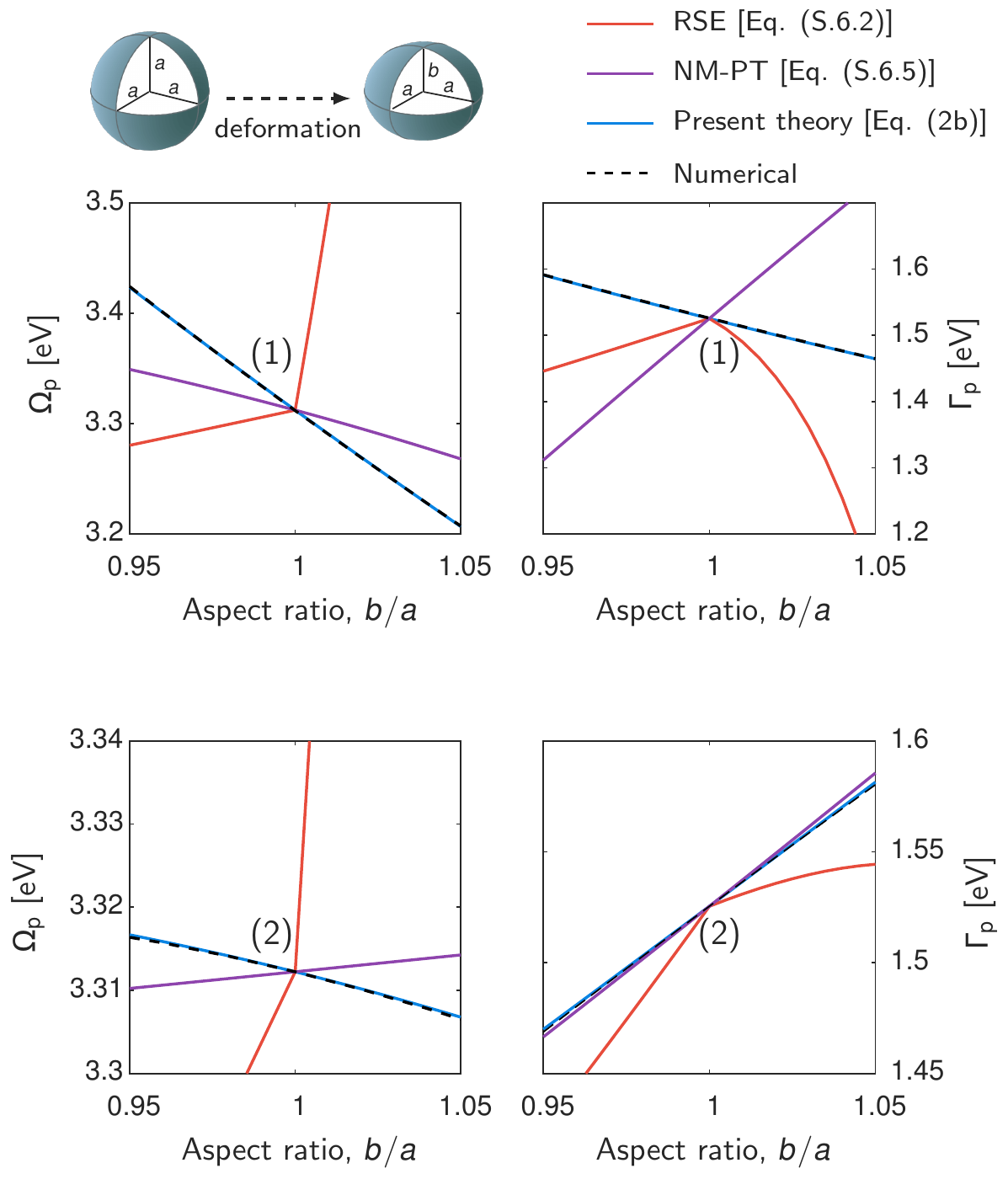}
\caption{
{\bf Comparisons of different PTs for predicting QNM frequencies} for a silver nanosphere deformed into spheroids.
The numerical example is a duplication of the left panel of Fig. 3b in the main text, wherein 30 QNMs are used in the PTs. Modal degeneracy factors are given in parenthesis.
}
\label{Fig:1stDiffC}
\end{figure}

\item \emph{Normal-mode} PT for deformation problems have also been developed in the Hermitian context by expanding the perturbed normal modes (NMs) with a series of the unperturbed NMs. In contrast with the previous example, the basis completeness is now guaranteed in all the “closed” space and thus the theory equally applies to inward or outward deformations. Hereafter, we refer to one of the most advanced NM PTs that thoroughly addresses the issue of local-field corrections~\cite{Harrington:1961} for deformation problems, by considering only field components that are continuous across $\partial V_{\rm res}$ ~\cite{Johnson:2002}\footnote{Note that, in Ref.~\cite{Johnson:2002}, only the first-order correction to a single mode is discussed, and we here generalize it to include multi-mode couplings with Eq. \eqref{eq:H1storder_OTH2}.}:
    \begin{align}
      \mathcal{H}_{{\rm p}; nm} \simeq
       \mel{\widetilde\Ev_{{\rm u};n}^{\scriptscriptstyle\parallel}}{
       h \Delta\varepsilon(\widetilde\omega_{{\rm u};m})}{\widetilde\Ev_{{\rm u};m}^{\scriptscriptstyle\parallel}}_{\scriptscriptstyle \partial V_{\rm res}}-
          \mel{\widetilde{\rm D}_{{\rm u};n}^{\scriptscriptstyle\perp}}{
       h \Delta\varepsilon^{-1}(\widetilde\omega_{{\rm u};m})}{\widetilde{\rm D}_{{\rm u};m}^{\scriptscriptstyle\perp}}_{\scriptscriptstyle \partial V_{\rm res}},
        \label{eq:H1storder_OTH2}
      \end{align}
      where $\widetilde\Ev_{{\rm u};n}^{\scriptscriptstyle\parallel}$ denotes (vectorial) parallel electric fields on $\partial V_{\rm res}$, while $\widetilde{\rm D}_{{\rm u};n}^{\scriptscriptstyle\perp}$ denotes (scalar) perpendicular electric displacement fields; $\Delta\varepsilon^{-1}\equiv\varepsilon_{\rm res}^{-1}-\varepsilon_{\rm bg}^{-1}$.

     Restricting the discussions to transverse modes\footnote{\label{note1}Note that the continuous-field method of Ref.~\cite{Johnson:2002} only applies for transverse modes with $\nablav\cdot \widetilde\Dv_{{\rm u};n}=0$ and $\widetilde\omega_{{\rm u};n}\ne 0$, and it becomes defective for zero-frequency (longitudinal) modes with $\nablav\times \widetilde\Ev_{{\rm u};n}=0$ and $\widetilde\omega_{{\rm u};n}=0$. Zero-frequency modes do not necessarily respect the conventional boundary conditions of transverse modes. Their perpendicular component of electric displacement fields and parallel components of electric fields can be discontinuous across the boundary, physically accounting for accumulated surface charges and dipole layers.}, and using the notation of the present article, the previous equation can be reformulated as
     \begin{align}
       \mathcal{H}_{{\rm p}; nm} \simeq
       \mel{\widetilde\Ev_{{\rm u};n}^{\rm bg}}{
       h \Delta\varepsilon(\widetilde\omega_{{\rm u};m})}{\widetilde\Ev_{{\rm u};m}^{\rm res}}_{\scriptscriptstyle \partial V_{\rm res}},
               \label{eq:H1storder_OTH3}
      \end{align}
      where the background medium is assumed to be dispersionless. Equation \eqref{eq:H1storder_OTH3} is similar with Eq. (2b) except that the overlap integral here involves the complex conjugated $\widetilde\Ev_{{\rm u};n}^{\rm bg}$, a signature of the Hermiticity.

We test the accuracy of the NM PT with the nanosphere case. As shown in Fig.~ \ref{Fig:1stDiffC}, the formalism (abbreviated as “NM-PT") cannot accurately predict first-order modal frequency shift of plasmonic resonators even for small geometrical deformation parameter $b/a \simeq 1$, consistently with earlier studies~\cite{Yang:2015}. Additionally, let us further mention that in earlier works on NM PTs, including in Ref.~\cite{Johnson:2002}, only first-order PT is derived and high-order PT is {impossible} to be derived~\footnote{See footnote 2.}.

       %The resulted first-order frequency shift $\Delta\widetilde\omega_{n}$ is thus given by
       %\begin{align}
       %\Delta\widetilde\omega_{n}= -\mel{\widetilde\Ev_{{\rm u};n}^{\rm bg}}{ h\Delta\bm\varepsilon(\widetilde\omega_{{\rm u};n})}{\widetilde\Ev_{{\rm u};n}^{\rm res}}_{\partial\Omega_u}.
       %\label{eq:PTSHITNM}
       %\end{align}
       %As latter evidenced numerically, Eq. \eqref{eq:PTSHITNM} cannot accurately predict perturbation-induced changes of QNM frequencies.
       %Moreover, we note that, in Ref.~\cite{Johnson:2002}, only the first-order PT is given, and the high-order PTs are difficult to be derived using the continuous-field method.

\end{enumerate}

The present work capitalizes on a series of earlier works on PTs of resonators. Essential ingredients towards unprecedented accurate predictions for large shape deformations are the embodiment of the extrapolation technique for representing electric fields in perturbed domains (implementing local field correction equivalently) with a relevant use of an advanced modal formalism combining dominant QNMs and additional numerical modes~\cite{Vial:2014}. The former allows us to obtain accurate predictions of large frequency changes with a low number of retained QNMs, as exemplified in Fig.~\ref{Fig:1stSiTikz} and Fig.~3 (main text), while the latter sets up a rigorous framework for handling both inward and outward deformations.

%-------------------------------------------------------------------------

\section{Lippman-Schwinger Formalism for QNMs}
\label{Sec:LSFormalism}
In this section, we derive the Lippman–Schwinger formalism for QNMs of perturbed systems, which expresses modal electric fields of the perturbed system using the Green's tensor of the unperturbed system.
In the perturbed system, electric fields of QNMs, $\widetilde\Ev_{\rm p}$, satisfy the source-free Maxwell's electric-field wave equation
\begin{subequations}
\begin{align}
\left[\nablav\times\mu_0^{-1}\nablav\times-\omega^2\bm\varepsilon_{\rm p}(\rv;\omega) \right]\widetilde\Ev_{\rm p}(\rv;\omega)=0.
\label{eq:E-p}
\end{align}
We reorganize Eq. \eqref{eq:E-p} into Eq. \eqref{eq:E-p2} manifesting the difference between the unperturbed system and the perturbed one:
\begin{align}
\left[\nablav\times\mu_0^{-1}\nablav\times-\omega^2\bm\varepsilon_{\rm u}(\rv;\omega) \right]\widetilde\Ev_{\rm p}(\rv;\omega)=
\omega^2\bm\left[\bm\varepsilon_{\rm p}(\rv;\omega)-\bm\varepsilon_{\rm u}(\rv;\omega)\right]\widetilde\Ev_{\rm p}(\rv;\omega).
\label{eq:E-p2}
\end{align}
\end{subequations}
The Green's tensor of the unperturbed system, $\greenOPF_{\rm u}$, satisfies
\begin{align}
\left[\nablav\times\mu_0^{-1}\nablav\times -\omega^2\bm\varepsilon_{\rm u}(\rv;\omega) \right]\greenOPF_{\rm u}(\rv,\rv';\omega)=\Iv\delta(\rv-\rv'),
\label{eq:E-green}
\end{align}
with second-rank identity tensor $\Iv$. Eq. \eqref{eq:E-green} together with \eqref{eq:E-p2} leads to
\begin{align}
\widetilde\Ev_{\rm p}(\rv;\omega)&=\omega^2\int \greenOPF_{\rm u}(\rv,\rv';\omega)\left[\bm\varepsilon_{\rm p}(\rv';\omega)-\bm\varepsilon_{\rm u}(\rv';\omega)\right]\widetilde\Ev_{\rm p}(\rv';\omega) d^3\rv'\nonumber\\
&=\omega^2\int \greenOPF_{\rm u}(\rv,\rv';\omega) f(\rv')\Delta\bm\varepsilon(\omega) \widetilde\Ev_{\rm p}(\rv';\omega) d^3\rv',
\label{eq:L-S}
\end{align}
where
$\bm\varepsilon_{\rm p}(\rv;\omega)-\bm\varepsilon_{\rm u}(\rv;\omega)= f(\rv)\left(\bm\varepsilon_{\rm res}(\rv;\omega)-\bm\varepsilon_{\rm bg}(\rv;\omega)\right)\equiv f(\rv)\Delta\bm\varepsilon(\omega)$ with filling function $f(r)$ that has values of $1$ and $-1$
for $\rv\in\mathbb{P}_{\rm res}$ and $\rv\in\mathbb{P}_{\rm bg}$, respectively, and a value of 0 elsewhere. Equation \eqref{eq:L-S} is the Lippman–Schwinger formalism for perturbed QNMs.

\section{Derivations of Eqs. (1)}
\label{Sec:DerEq1}

In this section, we derive Eqs. (1) in the main text. First, recall that $\widetilde\Ev_{\rm p}$ in $\bm {\mathbb{P}}_{\rm res, bg}$ domains, is extrapolated by the Taylor series expansion using $\widetilde\Ev_{\rm p}$ in $\bm {\mathbb{U}}_{\rm res, bg}$ domains:
\begin{align}
\widetilde\Ev_{\rm p}(\rv;\omega)=\sum_{j=0}^\infty \frac{l^j}{j!} {\overrightarrow{\partial}}_{\nv}^{j}\widetilde\Ev_{\rm p}(\rv_{\scriptscriptstyle \partial V_{\rm res}}+\sigma[h]\nv;\omega),
\label{eq:E-extrap}
\end{align}
where $\rv=\rvb+l\nv$ with $l\in[0,\,\,h]$, $\sigma[h]\equiv 0^+$ for $h<0$ and $\sigma[h]\equiv 0^-$ for $h>0$.
Eq. \eqref{eq:L-S}, together with Eq. \eqref{eq:E-extrap}, leads to Eqs. (1) through the following algebra manipulations.
\begin{align}
\widetilde\Ev_{\rm p}(\rv;\omega) &\overset{a}{=} \omega^2\Delta\bm\varepsilon(\omega) \sum_{j=0}^\infty  \int \greenOPF_{\rm u}(\rv,\rv';\omega) f(\rv')\frac{l^j}{j!}  {\overrightarrow{\partial}}_{\nv}^{j}\widetilde\Ev_{\rm p}(\rv_{\scriptscriptstyle \partial V_{\rm res}}'+\delta[h]\nv;\omega)d^3\rv'\nonumber\\
&\overset{b}{=} \omega^2\Delta\bm\varepsilon(\omega)\sum_{k=0}^\infty \sum_{j=0}^\infty   \int  \greenOPF_{\rm u}(\rv,\rv_{\scriptscriptstyle \partial V_{\rm res}}'-\delta[h]\nv;\omega)  {\overleftarrow{\partial}}_{\nv}^k
f(\rv')\frac{l^{k+j}}{k!j!}
{\overrightarrow{\partial}}_{\nv}^{j}\widetilde\Ev_{\rm p}(\rv_{\scriptscriptstyle \partial V_{\rm res}}'+\delta[h]\nv;\omega) d^3\rv'\nonumber\\
&\overset{c}{=} \omega^2 \Delta\bm\varepsilon(\omega)\sum_{k=0}^\infty \sum_{j=0}^\infty  \oint_{\partial V_{\rm res}}\int_0^{h}  \greenOPF_{\rm u}(\rv,\rv_{\scriptscriptstyle \partial V_{\rm res}}'{-}\delta[h]\nv;\omega)  {\overleftarrow{\partial}}_{\nv}^k
\frac{l^{k+j}}{k!j!} \left[1+l \kappa_1 (\rv_{\scriptscriptstyle \partial V_{\rm res}}')\right]
\left[1+l \kappa_2 (\rv_{\scriptscriptstyle \partial V_{\rm res}}') \right]\nonumber\\
&\quad\quad\quad\quad\quad\quad\quad\quad\quad\quad\quad\quad
{\overrightarrow{\partial}}_{\nv}^{j}\widetilde\Ev_{\rm p}(\rv_{\scriptscriptstyle \partial V_{\rm res}}'{+}\delta[h]\nv;\omega)
d^2\rv_{\scriptscriptstyle \partial V_{\rm res}}' dl \nonumber\\
&\overset{d}{=}  \omega^2 \Delta\bm\varepsilon(\omega) \sum_{k=0}^\infty \sum_{j=0}^\infty \oint_{\partial V_{\rm res}} \greenOPF_{\rm u}(\rv,\rv_{\scriptscriptstyle \partial V_{\rm res}}'-\delta[h]\nv;\omega)  {\overleftarrow{\partial}}_{\nv}^k
\frac{h^{j+k+1}(\rv_{\scriptscriptstyle \partial V_{\rm res}}')}{k!j!}\nonumber\\
&\quad\quad\quad\quad\quad\quad\quad\quad\quad\quad\left[\frac{1}{k+j+1}+\kappa_{\rm m}(\rv_{\scriptscriptstyle \partial V_{\rm res}}')\frac{2h(\rv_{\scriptscriptstyle \partial V_{\rm res}}')}{k+j+2}+\kappa_{\rm g}(\rv_{\scriptscriptstyle \partial V_{\rm res}}')\frac{h^2(\rv_{\scriptscriptstyle \partial V_{\rm res}}')}{k+j+3}\right]
{\overrightarrow{\partial}}_{\nv}^{j}\widetilde\Ev_{\rm p}(\rv_{\scriptscriptstyle \partial V_{\rm res}}'+\delta[h]\nv)
d^2\rv_{\scriptscriptstyle \partial V_{\rm res}}'\nonumber\\
&\overset{e}{=} \omega^2 \oint_{\partial V_{\rm res}} \greenOPF_{\rm u}(\rv,\rv_{\scriptscriptstyle \partial V_{\rm res}}-\delta[h]\nv;\omega)  \widetilde {\mathbf P}_{\rm Geom}(\rv_{\scriptscriptstyle \partial V_{\rm res}};\omega) d^2\rvb
\label{eq:Eq1SI}
\end{align}
The above derivations involve the following steps:
\begin{enumerate}[label=$\alph*$.\ ] \setlength{\parskip}{0pt}   \setlength{\itemsep}{2pt}
\item Insert Eq. \eqref{eq:E-extrap} into Eq. \eqref{eq:L-S}.
\item Employ the Taylor series expansion of $\greenOPF_{\rm u}(\rv,\rv';\omega)$ with $\rv'\in\mathbb{P}_{\rm res, bg}$ and $\rv\in\mathbb{U}_{\rm res, bg}$:
\begin{align}
\greenOPF_{\rm u}(\rv,\rv';\omega)&=\sum_{k=0}^\infty  \left[\frac{\partial\greenOPF_{\rm u}(\rv,\rv_{\scriptscriptstyle \partial V_{\rm res}}'-\delta\nv;\omega)}{\partial x'}n_x(\rv_{\scriptscriptstyle \partial V_{\rm res}}')+\frac{\partial\greenOPF_{\rm u}(\rv,\rv_{\scriptscriptstyle \partial V_{\rm res}}'-\delta\nv;\omega)}{\partial y'}n_y(\rv_{\scriptscriptstyle \partial V_{\rm res}}')+\frac{\partial\greenOPF_{\rm u}(\rv,\rv_{\scriptscriptstyle \partial V_{\rm res}}'-\delta\nv;\omega)}{\partial z'}n_z(\rv_{\scriptscriptstyle \partial V_{\rm res}}')\right]^k \frac{l^{k}}{k!}\nonumber\\
& \equiv \sum_{k=0}^\infty  \greenOPF_{\rm u}(\rv,\rv_{\scriptscriptstyle \partial V_{\rm res}}'-\delta\nv;\omega)  {\overleftarrow{\partial}}_{\nv}^k \frac{l^{k}}{k!},
\nonumber
\end{align}
where $n_{x,y,z}$ denote the three Cartesian components of $\nv$. Note that the Taylor series expansion of $\greenOPF_{\rm u}(\rv,\rv';\omega)$ is meaningful here because $\rv$ and $\rv'$ are in different domains ($|\rv-\rv'|\ne0$) ensuring the regularity of $\greenOPF_{\rm u}$.
\item  (1) Application of the identity $d^3\rv'=  \left[1+l \kappa_1 (\rv_{\scriptscriptstyle \partial V_{\rm res}}') \right]\left[1+l \kappa_2 (\rv_{\scriptscriptstyle \partial V_{\rm res}}') \right]d^2\rvb'dl$, where $\kappa_{1,2}$ denotes the two principal curvatures of $\partial V_{\rm res}$. (2) The filling function $f$ is implicit in $\int_0^ h\cdots dl$.
\item Integrate out the variable $l$, and introduce the mean curvature $\kappa_{\rm m}\equiv\kappa_1/2+\kappa_2/2$ and the Gaussian curvature $\kappa_{\rm g}\equiv\kappa_1\kappa_2$.
\item Application of the definition of $\widetilde {\mathbf P}_{\rm Geom}$ [Eq. (1b) in the main text].
\end{enumerate}

\section{Perturbation Theory}

Deriving the modal perturbation theory (PT) departs from the following equations [Eqs. (1) in the main text]

\begin{subequations}
\begin{align}
\widetilde\Ev_{\rm p}(\rv;\omega)= \omega^2\oint_{\partial V_{\rm res}} \greenOPF_{\rm u}(\rv,\rv_{\scriptscriptstyle \partial V_{\rm res}}'-\delta[h]\nv;\omega) \widetilde{\mathbf P}_{\rm Geom}(\rv_{\scriptscriptstyle \partial V_{\rm res}}';\omega) d^2\rvb',
\label{eq:MT1}
\end{align}
with
\begin{align}
\widetilde{\mathbf P}_{\rm Geom}(\rv_{\scriptscriptstyle \partial V_{\rm res}};\omega)  & =  \sum_{j=0}^{\infty}\sum_{k=0}^{\infty} {\overleftarrow{\partial}}_{\nv}^k
{\rm c}_{jk}(\rvb)
 \Delta\bm\varepsilon (\omega )
{\overrightarrow{\partial}}_{\nv}^{j}\widetilde\Ev_{\rm p}(\rv_{\scriptscriptstyle
\partial V_{\rm res}}+\delta[h]\nv;\omega),
\label{eq:PGEOM}
\end{align}
and
\begin{align}
{\rm c}_{jk}(\rvb)=\frac{h(\rvb)^{k+j+1}}{k!j!}\left(
\frac{1}{k+j+1}+\kappa_{\rm m}(\rvb)\frac{2h(\rvb)}{k+j+2}+
\kappa_{\rm g}(\rvb)\frac{h(\rvb)^2}{k+j+3}\right).
\end{align}
\end{subequations}

%First, we multiply $\Delta\bm\varepsilon\equiv \bm \varepsilon_{\rm res}- \varepsilon_{\rm bg}$ (the difference between resonator and background permittivity) on both sides of Eqs. \eqref{eq:MT1} and obtain that
%\begin{align}
%\Delta\bm\varepsilon (\omega) \Ev(\rv;\omega)= \iint_{\partial V_{\rm res}} \Delta\bm\varepsilon (\omega) \greenOPF_{\rm u}(\rv,\rv_{\scriptscriptstyle \partial V_{\rm res}}'-\sigma[h]\nv;\omega)  {\mathbf P}_{\rm Geom}(\rv_{\scriptscriptstyle \partial V_{\rm res}}'+\sigma[h]\nv;\omega) da.
%\label{eq:MT1}
%\end{align}

\subsection{Modal Expansions}
\label{Sec:QNMEXP}
Equation \eqref{eq:MT1} defines an eigenvalue equation for perturbed eigenmodes. However, this equation is nonlinear with respect to $\omega$, and is, thus, computationally difficult. To simplify the problem, we linearize Eq. \eqref{eq:MT1} exploiting the modal-expansion technique.

Moreover, since deformation problems often involve material changes outside nanoresonators, wherein QNMs alone cannot form a complete basis, we here incorporate extra numerical modes\footnote{Numerical modes are either due to space truncation by perfectly matched layers (PMLs) or numerical error from discretizing Maxwell's operator.}into our modal basis~\cite{Vial:2014,Yan:2018}. Therefore, the employed modal basis consists of a restricted set of dominant QNMs with additional numerical modes. Such a modal basis is complete over the whole space, which, thus, theoretically justify the exactness of the modal expansions employed below. Moreover, as evidenced with numerical examples, employing a few dominant QNMs already delivers quite accurate results. Thus, numerical modes need not be used in practical simulations, and they are introduced just for the rigorousness of the theory.

$\greenOPF_{\rm u}$ can be expanded as follows~\cite{Lalanne:2018}:
\begin{subequations}
\begin{align}
\greenOPF_{\rm u}(\rv,\rvb'-\delta[h]\nv;\omega)
=-\sum_n  \frac{ \widetilde\Ev_{{\rm u};n}(\rv)\otimes \widetilde\Ev_{{\rm u};n}(\rvb'-\delta[h]\nv)}{\omega(\omega-\widetilde\omega_{{\rm u};n})}. \label{eq:GQNMEXP}
\end{align}
Plugging Eq. \eqref{eq:GQNMEXP} into Eq. \eqref{eq:MT1} gives us
\begin{align}
\widetilde\Ev_{\rm p}(\rv;\omega)= \sum_ n \underbrace{\left[{- \frac{\omega}{\omega-\widetilde\omega_{{\rm u}; n}} \oint_{\partial V_{\rm res}} \widetilde\Ev_{{\rm u};n}(\rvb-\delta[h]\nv) \widetilde{\mathbf P}_{\rm Geom}(\rv_{\scriptscriptstyle \partial V_{\rm res}};\omega) d^2\rvb}\right]}_{\text{\normalsize $\equiv\alpha_n(\omega)$}}\widetilde\Ev_{{\rm u};n}(\rv),\nonumber
\end{align}
which suggests the expansion of $\widetilde\Ev_{\rm p}$ with
\begin{align}
\widetilde\Ev_{\rm p}(\rv;\omega)=\sum_{n}\alpha_{n}(\omega)\widetilde\Ev_{{\rm u};n}(\rv).
\end{align}
Further, we can prove that $\widetilde {\mathbf P}_{\rm Geom}$ has the same expansion coefficients $\alpha_{n}$ as $\widetilde\Ev_{\rm p}$ (c.f. the derivations at the end of this subsection) with
\begin{align}
\widetilde {\mathbf P}_{\rm Geom}(\rvb;\omega)  & =  \sum_n \sum_{j=0}^{\infty}\sum_{k=0}^{\infty} \alpha_{n} (\omega) {\overleftarrow{\partial}}_{\nv}^k
c_{jk}(\rvb) {\overrightarrow{\partial}}_{\nv}^{j} \Delta\bm\varepsilon (\widetilde\omega_{{\rm u};n} ) \widetilde\Ev_{{\rm u};n}(\rvb+\delta[h]\nv).
\label{eq:QNMEXPPGEOM}
\end{align}
\label{eq:QNMEXP}
\end{subequations}%

%As documented in the literatures~\cite{Lalanne:2018,Muljarov:2016,Muljarov:2016b}, $\greenOPF_{\rm u}$ has the following modal-expansion expression
%\begin{align}
%\greenOPF_{\rm u}(\rv,\rv';\omega)
% & =-\sum_n  \frac{ \widetilde\Ev_{{\rm u};n}(\rv)\otimes \widetilde\Ev_{{\rm u};n}(\rv')}{(\omega-\widetilde\omega_{{\rm u};n})\omega}.
%\end{align}
%The expansion is
%QNMs are a rigorous description of resonance states in open systems. Mathematically, they are poles of the Green's tensor, $\greenOPF_{\rm u}$ defined in Eq. \ref{eq:E-green}. The poles of the Green's tensor can be identified into two types: (1) one at zero frequency; (2) the other at non-zero frequency. The non-zero-frequency poles are relevant with resonance phenomena observed in experiment and they are thus physical modes, while the zero-frequency poles, corresponding to the so-called static modes, are less studied except in a series of papers by E. A. Muljarov and W. Langbein. The documented literatures suggest a few
%different expressions of $\greenOPF_{\rm u}$ (see Ref. [xx] for a summary); here we list two expressions:
%\begin{align}
%\end{align}
Plugging Eqs. \eqref{eq:QNMEXP} into Eq. \eqref{eq:MT1}, defining $\ket{\bm\alpha} \equiv \left[\alpha_1;\alpha_2;\cdots; \alpha_n; \cdots \right]$ and introducing $\widetilde\omega_{\rm p}$ to denote frequencies of perturbed modes, we arrive at
\begin{subequations}
\begin{align}
{\bf{\mathcal{H}}}_0 \ket{\bm\alpha}=
\widetilde\omega_{\rm p}\left(\Iv +\bf{\mathcal{H}}_{\rm p} \right)\ket{\bm\alpha},
\label{eq:SIPTa}
\end{align}
with
\begin{align}
{\bf \mathcal{H}}_{0; nm}=\widetilde\omega_{{\rm u};n} \delta_{nm}
\label{eq:SIPT0}
\end{align}
and
\begin{align}
\mathcal{H}_{{\rm p};nm}
&=
\sum_{j=0}^{\infty}\sum_{k=0}^{\infty}
\oint_{\scriptscriptstyle\partial V_{\rm res}}
{\widetilde\Ev_{{\rm u};n}(\rvb-\delta[h]\nv)}\cdot
{ {\overleftarrow{\partial}}_{\nv}^k c_{jk}(\rvb)\Delta\bm\varepsilon(\widetilde\omega_{{\rm u};m})} {\overrightarrow{\partial}}_{\nv}^{j}{\widetilde\Ev_{{\rm u};m}(\rvb+\delta[h]\nv)}\,d^2\rvb\nonumber\\
&\equiv
\sum_{j=0}^{\infty}\sum_{k=0}^{\infty}
\mel{\left(\widetilde\Ev_{{\rm u};n}^{-\delta[h]}\right)^*}{ {\overleftarrow{\partial}}_{\nv}^k
c_{jk} \Delta\bm\varepsilon(\widetilde\omega_{{\rm u};m}) {\overrightarrow{\partial}}_{\nv}^{j}}{\widetilde\Ev_{{\rm u};m}^{\delta[h]}}_{\scriptscriptstyle \partial V_{\rm res}},
\label{eq:SIPTb}
\end{align}
where, in the second line, a shorthand Dirac notation is introduced for the overlap integral.
Equations \eqref{eq:SIPT} define a linear eigenvalue equation for perturbed modes.
\label{eq:SIPT}
\end{subequations}

\subsubsection*{\bf  Proof of Eq. \eqref{eq:QNMEXPPGEOM}}

Multiplying $\sum_{j=0}^{\infty}\sum_{k=0}^{\infty} {\overleftarrow{\partial}}_{\nv}^k
c_{jk}(\rvb) \Delta\bm\varepsilon (\omega ) {\overrightarrow{\partial}}_{\nv}^{j} $ on both sides of Eq. \eqref{eq:MT1}, we arrive at
\begin{align}
\widetilde{\mathbf P}_{\rm Geom}(\rvb;\omega)= \omega^2\sum_{j=0}^{\infty}\sum_{k=0}^{\infty} {\overleftarrow{\partial}}_{\nv}^k
c_{jk}(\rvb)\Delta\bm\varepsilon (\omega )  {\overrightarrow{\partial}}_{\nv}^{j} \oint_{\partial V_{\rm res}} \greenOPF_{\rm u}(\rvb+\delta[h]\nv,\rvb'-\delta[h]\nv;\omega) \widetilde{\mathbf P}_{\rm Geom}(\rvb';\omega) d^2\rvb'.
\label{eq:MT2}
\end{align}
Then, employing the Mittag-Leffler theorem, we derive the modal expansion expression of $\Delta\bm\varepsilon (\omega ) \greenOPF_{\rm u}(\rvb-\delta[h]\nv,\rvb'+\delta[h]\nv;\omega)$ by identifying its poles and residues:
\begin{align}
\Delta\bm\varepsilon (\omega ) \greenOPF_{\rm u}(\rvb+\delta[h]\nv,\rvb'-\delta[h]\nv;\omega)=
-\sum_n\frac{\Delta\bm\varepsilon(\widetilde\omega_{{\rm u}; n})\widetilde\Ev_{{\rm u}; n}(\rvb+\delta[h]\nv)\otimes\widetilde\Ev_{{\rm u}; n}(\rvb'-\delta[h]\nv)}{\omega(\omega-\widetilde\omega_{{\rm u}; n})}.
\label{eq:DGQNM}
\end{align}
In particular, when $\Delta\bm\varepsilon$ has a pole at zero frequency~\footnote{$\Delta\bm\varepsilon(\omega)\propto 1/\omega \to \infty$ as $\omega\to 0$, e.g., for Drude metals.}, the numerator term in Eq. \eqref{eq:DGQNM} for zero-frequency modes ($\widetilde\omega_{{\rm u}; n}=0$), represents a limit:
$\lim_{\Delta\bm\varepsilon(0)\to\infty}\Delta\bm\varepsilon(0)\widetilde\Ev_{{\rm u}; n}(\rvb+\delta[h]\nv)\otimes\widetilde\Ev_{{\rm u}; n}(\rvb'-\delta[h]\nv)$. While for non-zero poles of $\Delta\bm\varepsilon$, they do not need to be included in Eq. \eqref{eq:DGQNM} as explained below.
\begin{itemize}
 \item Consider that $\Delta\bm\varepsilon$ has a non-zero pole at $\Omega_j$. First, we discuss a specific case: $\bm\varepsilon_{\rm res}\propto 1/(\omega-\Omega_j)\to\infty$ as $\omega\to\Omega_j$, while $\bm\varepsilon_{\rm bg}$ is regular; $\delta[h]=0^+$ due to $h<0$, and, thus, $\rvb-\delta[h] \nv$ and $\rvb'+\delta[h] \nv$ represent the inner (resonator) and outer (background) sides of $\partial V_{\rm res}$, respectively. In this case, $\greenOPF_{\rm u}(\rvb+\delta[h]\nv,\rvb'-\delta[h]\nv;\omega)$ associates with electric fields at $\rvb+\delta[h]\nv$ generated by a point dipole at $\rvb'-\delta[h]\nv$. Obviously, $\greenOPF_{\rm u}(\rvb+\delta[h]\nv,\rvb'-\delta[h]\nv;\omega)\to 0$ as $\omega\to\Omega_j$, since no fields can transmit into a medium with an infinitely large permittivity. The vanishing $\greenOPF_{\rm u}$ compensates with the diverging $\Delta\bm\varepsilon$, thereby eliminating the material pole at $\Omega_j$. If $\delta[h]=0^-$ (due to $h>0$), the reciprocity of the Green's tensor ensures the validity of the same conclusion. Moreover, for the case $\bm\varepsilon_{\rm bg}\to \infty$ as $\omega\to\Omega_j$, following the same routines, we again derive that $\Delta\bm\varepsilon (\omega )  \greenOPF_{\rm u}(\rvb+\delta[h]\nv,\rvb'-\delta[h]\nv;\omega)$ has no pole at $\Omega_j$.

     \emph{In summary, non-zero frequency poles of $\Delta\bm\varepsilon$ are not poles to
     $\Delta\bm\varepsilon (\omega ) \greenOPF_{\rm u}(\rvb+\delta[h]\nv,\rvb'-\delta[h]\nv;\omega)$.}
\end{itemize}

Substituting Eq. \eqref{eq:DGQNM} into Eq. \eqref{eq:MT2}, we derive that
\begin{align}
\widetilde{\mathbf P}_{\rm Geom}(\rvb;\omega)&=\sum_n\sum_{j=0}^{\infty}\sum_{k=0}^{\infty}\underbrace{\left[-\frac{\omega}{\omega-\widetilde\omega_{{\rm u}; n}}
\oint_{\partial V_{\rm res}} \widetilde\Ev_{{\rm u};n}(\rvb'-\delta[h]\nv) \widetilde{\mathbf P}_{\rm Geom}(\rv_{\scriptscriptstyle \partial V_{\rm res}}';\omega) d^2\rvb'\right]}_{ \text{\normalsize =$\alpha_n(\omega)$}}
\nonumber\\
&\quad\quad\quad\quad\quad\quad{\overleftarrow{\partial}}_{\nv}^k
c_{jk}(\rvb)\Delta\bm\varepsilon(\widetilde\omega_{{\rm u};n}) {\overrightarrow{\partial}}_{\nv}^{j}\widetilde\Ev_{{\rm u}; n}(\rvb+\delta[h]\nv) \nonumber\\
&=\sum_n\sum_{j=0}^{\infty}\sum_{k=0}^{\infty}
\alpha_n({\omega})
{\overleftarrow{\partial}}_{\nv}^k
c_{jk}(\rvb) \Delta\bm\varepsilon(\widetilde\omega_{{\rm u};n}) {\overrightarrow{\partial}}_{\nv}^{j}\widetilde\Ev_{{\rm u}; n}(\rvb+\delta[h]\nv),
\end{align}
thereby concluding the proof of Eq. \eqref{eq:QNMEXPPGEOM}.

%We assume that the permittivity tensors of the resonator and the background are expressed as
%\begin{align}
%\bm\varepsilon_{\scriptscriptstyle \rm Mat}(
%\omega)=\bm\varepsilon_{{\scriptscriptstyle \rm Mat};\infty}+\sum_j \frac{\bm w_{{\scriptscriptstyle \rm Mat};j}}{\omega-\Omega_j} +
%\frac{-\bm w_{{\scriptscriptstyle \rm Mat};j}^* (\rv)}{\omega+\Omega_j^*}.
%\end{align}
%Here $\rm Mat=\left\{res,\;bg\right\}$; $\varepsilon_{{\scriptscriptstyle \rm Mat};\infty}$ and $\bm w_{{\scriptscriptstyle \rm Mat};j}$ are symmetric tensors since we only consider reciprocity materials in the present paper; the particular setting of material poles, appearing in pairs---$\Omega_j$ and $-\Omega_j^*$---with strength tensors $\bm w_{{\scriptscriptstyle \rm Mat};j}$ and $-\bm w_{{\scriptscriptstyle \rm Mat};j}^*$, respectively, is due to the causality principle that requires $\bm\varepsilon_{\scriptscriptstyle \rm Mat}(\rv;\omega)^*=\bm\varepsilon_{\scriptscriptstyle \rm Mat}(\rv;-\omega^*)$. The permittivity difference between the resonator and background are due to the different values between $\bm\varepsilon_{\scriptscriptstyle \rm res}$ and $\bm\varepsilon_{\scriptscriptstyle \rm bg}$, between
%$\bm w_{{\scriptscriptstyle \rm res};j}$ and $\bm w_{{\scriptscriptstyle \rm bg};j}$.

\subsection{ First-order Perturbation Theory}
\label{Sec:1stQP_Theory}

By retaining the leading term $j=k=0$ in Eq. \eqref{eq:SIPTb}, we get the first-order approximation of $\mathcal{H}_{\rm p}$:
\begin{align}
\mathcal{H}_{{\rm p}; nm} \simeq
\mel{\left(\widetilde\Ev_{{\rm u};n}^{-\sigma[h]}\right)^*}{
h \Delta\bm\varepsilon(\widetilde\omega_{{\rm u};m})}{\widetilde\Ev_{{\rm u};m}^{\sigma[h]}}_{\scriptscriptstyle \partial V_{\rm res}}.
\label{eq:H1storderb}
\end{align}
Combining Eq. \eqref{eq:H1storderb} with Eq. \eqref{eq:SIPTa} gives us the first-order PT.

A correct PT should ensure that perturbed eigensolutions are analytic with respect to perturbation parameter.
%Thereupon, the proposed first-order QNM PT should at least lead to that both $\left.\delta\widetilde\omega_{\rm p}/\delta h\right|_{h=0}$ and $\left.\delta \ket{\bm\alpha}/\delta h\right|_{h=0}$ (the first-order functional derivative of perturbed frequencies and eigenstates with respect to $h$ at the non-perturbed point $h=0$) are well defined mathematically.
This requirement, however, is not transparent with Eq. \eqref{eq:H1storderb}, since the latter completely changes its expression as $h$ changes its sign. To evidence the analyticity of the first-order QNM PT, we better take an alternative, equivalent expression (see the derivations at the end of this subsection)
\begin{align}
\tcboxmath
{
\mathcal{H}_{{\rm p}; nm}\simeq
\mel{\left(\widetilde\Ev_{{\rm u};n}^{\rm res}\right)^*}{
h \Delta\bm\varepsilon(\widetilde\omega_{{\rm u};m})}{\widetilde\Ev_{{\rm u};m}^{\rm bg}}_{\scriptscriptstyle \partial V_{\rm res}},}
\label{eq:H1storder}
\end{align}
which is Eq. (2b) in the main text. Here, $\widetilde\Ev_{{\rm u};n}^{\rm bg}\equiv \widetilde\Ev_{{\rm u};n}(\rvb+0^+\nv)$ and
$\widetilde\Ev_{{\rm u};n}^{\rm res}\equiv \widetilde\Ev_{{\rm u};n}(\rvb+0^-\nv)$ denote $\widetilde\Ev_{{\rm u};n}$ at the outer and inner sides of
$\partial V_{\rm res}$, respectively. The analyticity of Eqs. \eqref{eq:H1storder} with respect to $h$ thus ensures that $\widetilde\omega_{\rm p}$ and $ \ket{\bm\alpha}$ are also analytic.

\subsubsection*{\bf Proof of equivalence between Eqs. \eqref{eq:H1storderb} and \eqref{eq:H1storder}}

To prove the equivalence between Eqs. \eqref{eq:H1storderb} and \eqref{eq:H1storder}, we retreat to Eq. \eqref{eq:MT1}, wherein the first-order approximation of $\widetilde{\mathbf P}_{\rm Geom}$ is given by
\begin{align}
\widetilde{\mathbf P}_{\rm Geom}(\rvb)\simeq\Delta \bm \varepsilon (\omega) h \widetilde\Ev_{\rm p}(\rvb+\delta\nv;\omega) . \nonumber
\end{align}
$\widetilde{\mathbf P}_{\rm Geom}$ physically represents a surface polarization, which, depending on the sign of $h$, locates either at the inner or outer sides of $\partial V_{\rm res}$. It is this sign dependence that hides the analyticity of the first-order PT.
To eliminate the sign dependence, we move the part of $\widetilde{\mathbf P}_{\rm Geom}$ at the inner side of $\partial V_{\rm res}$ to the outer side, while requiring radiated fields intact. This is achieved by changing the expression of $\widetilde{\mathbf P}_{\rm Geom}$ for $h<0$ with
\begin{align}
\widetilde{\mathbf P}_{\rm Geom}^{\rm C}(\rvb;\omega)=
\left\{
                \begin{array}{ll}
                 \Delta \bm \varepsilon (\omega) h \widetilde\Ev_{\rm p}(\rvb+0^-\nv;\omega)    & \quad h>0, \\
                 {\left[{\bf\mathcal{M}}_{\rm bg-res}^{\rm T}\right]^{-1}}\Delta \bm \varepsilon (\omega) h  \widetilde\Ev_{\rm p}(\rvb+0^+\nv;\omega)     &\quad h< 0.
                \end{array}
\right.
\label{eq:PGEOMC}
\end{align}
Here ${\bf\mathcal{M}}_{\rm bg-res}$ is a $3 \times 3$ matrix defined by
\begin{align}
\text{${\bf\mathcal{M}}_{\rm bg-res}= {\bf\mathcal{M}}_{\rm f-bg}^{-1}{\bf\mathcal{M}}_{\rm f-res}$, with
${\bf\mathcal{M}}_{\rm f-\eta}=(\Iv-\nv\nv)+\nv\nv\bm\varepsilon_{\eta}$ and $\eta=\left\{\rm bg, res\right\}$}.
\end{align}
${\bf\mathcal{M}}_{\rm bg-res}$ relates the electric fields at the inner and outer sides of $\partial V_{\rm res}$--- $\Ev(\rvb+0^+\nv)={\bf\mathcal{M}}_{\rm bg-res}\Ev(\rvb+0^-\nv)$---by the traditional boundary conditions: Parallel electric fields and perpendicular displacement electric fields keep continuous across the boundary of different media.

Even though its physical position is now independent of the sign of $h$, $\widetilde{\mathbf P}_{\rm Geom}^{\rm C}$ still remains the sign dependence in its expression, c.f. Eq. \eqref{eq:PGEOMC}. To further remove this dependence, we express $\widetilde\Ev_{\rm p}(\rvb+0^+\nv)$ in terms of  $\widetilde\Ev_{\rm p}(\rvb+0^-\nv)$ with the boundary conditions that take the surface polarization $\widetilde{\mathbf P}_{\rm Geom}^{\rm C}$ into account, and arrive at
\begin{align}
\widetilde\Ev_{\rm p}(\rvb+0^+\nv;\omega)=
{\bf\mathcal{M}}_{\rm bg-res}
\widetilde\Ev_{\rm p}(\rvb+0^-\nv;\omega) -
\underbrace{{\bf\mathcal{M}}_{\rm f-bg}^{-1}{\bf \mathcal{K}} h \widetilde\Ev_{\rm p}(\rvb+0^+\nv;\omega)}_{\text{\normalsize contribution from the surface polarization}},
\label{eq:Edurel}
\end{align}
with
\begin{align}
{\bf \mathcal{K}} h \widetilde\Ev_{\rm p}(\rvb+0^+\nv;\omega)\equiv -\bm\nabla_{\scriptscriptstyle\parallel}&\left[({\bf\mathcal{M}}_{\rm f-res}^{\rm T})^{-1}\Delta\bm\varepsilon\widetilde\Ev_{\rm p}(\rvb+0^+\nv;\omega) h\cdot\nv \right]\nonumber\\
-\bm\nabla_{\scriptscriptstyle\parallel}\cdot&\left[({\bf\mathcal{M}}_{\rm f-res}^{\rm T})^{-1}\Delta\bm\varepsilon\widetilde\Ev_{\rm p}(\rvb+0^+\nv;\omega) h\cdot (\Iv-\nv\nv)
\right].
\end{align}
Plugging Eq. \eqref{eq:Edurel} into Eq. \eqref{eq:PGEOMC} and employing the identity
\footnote{{Proof of $\left({\bf\mathcal{M}}_{\rm bg-res}^{\rm T}\right)^{-1}\Delta\bm\varepsilon {\bf\mathcal{M}}_{\rm bg-res} = \Delta\bm\varepsilon$}:\\
At $\partial V_{\rm res}$, we choose a local Cartesian coordinate system $\left\{x,y,z\right\}$ such that $x$ and $y$ are coordinates that are   parallel with $\partial V_{\rm res}$, while $z$ is along the perpendicular direction of $\partial V_{\rm res}$.
Employing this coordinate system, ${\bf\mathcal{M}}_{\rm f-res}$ and ${\bf\mathcal{M}}_{\rm f-bg}$ are expressed as
\begin{align}
{\bf\mathcal{M}}_{\rm f-res}=
\begin{bmatrix}
1 & 0 & 0 \\
0 & 1 & 0 \\
\varepsilon_{{\rm res}; zx}  & \varepsilon_{{\rm res}; zy} & \varepsilon_{{\rm res}; zz}
\end{bmatrix},\quad\quad
{\bf\mathcal{M}}_{\rm f-bg}=
\begin{bmatrix}
1 & 0 & 0 \\
0 & 1 & 0 \\
\varepsilon_{{\rm bg}; zx}  & \varepsilon_{{\rm bg}; zy} & \varepsilon_{{\rm bg}; zz}
\end{bmatrix}.
\nonumber
\end{align}
Then, with the relation ${\bf\mathcal{M}}_{\rm bg-res}={\bf\mathcal{M}}_{\rm f-bg}^{-1}{\bf\mathcal{M}}_{\rm f-res}$, we derive that
\begin{align}
{\bf\mathcal{M}}_{\rm bg-res}=\begin{bmatrix}
1 & 0 & 0 \\
0 & 1 & 0 \\
\frac{\varepsilon_{{\rm res}; zx}- \varepsilon_{{\rm bg}; zx}}{\varepsilon_{{\rm bg}; zz}} &\frac{\varepsilon_{{\rm res}; zy}- \varepsilon_{{\rm bg}; zy}}{\varepsilon_{{\rm bg}; zz}} & \frac{\varepsilon_{{\rm res}; zz}}{\varepsilon_{{\rm bg}; zz}}
\end{bmatrix}
.
\nonumber
\end{align}
For reciprocal materials with $\bm \varepsilon_{\rm bg, res}=\bm \varepsilon_{\rm bg, res}^{\rm T}$, it is straightforward to confirm that
\begin{align}
\Delta\bm\varepsilon {\bf\mathcal{M}}_{\rm bg-res} = {\bf\mathcal{M}}_{\rm bg-res}^{\rm T} \Delta\bm\varepsilon.
\nonumber
\end{align}
The above equation is then reformulated as $\left({\bf\mathcal{M}}_{\rm bg-res}^{\rm T}\right)^{-1}\Delta\bm\varepsilon {\bf\mathcal{M}}_{\rm bg-res} = \Delta\bm\varepsilon$.
}
\begin{align}
\left({\bf\mathcal{M}}_{\rm bg-res}^{\rm T}\right)^{-1}\Delta\bm\varepsilon {\bf\mathcal{M}}_{\rm bg-res} = \Delta\bm\varepsilon \quad\text{for reciprocal materials with $\Delta\bm\varepsilon=\Delta\bm\varepsilon^{\rm T}$}\nonumber,
\end{align}
we obtain
\begin{align}
\widetilde{\mathbf P}_{\rm Geom}^{\rm C}(\rvb;\omega)
=\underbrace{\Delta \bm \varepsilon (\omega) h \widetilde\Ev_{\rm p}(\rvb+0^-\nv;\omega) }_{\text{\normalsize$\propto  h$}}
                 -
                 \underbrace
                 {
                \Delta \bm \varepsilon (\omega) h {\bf\mathcal{M}}_{\rm f-res}^{-1} {\bf \mathcal{K}} h \widetilde\Ev_{\rm p}(\rv+0^+\nv;\omega)
                \theta[-h]
                }_
                {\text{\normalsize $\propto h^2$}},
\label{eq:PGEOMC2}
\end{align}
with the Heaviside step function $\theta(x)$ having $\theta(x)=1$ for $x>0$ and otherwise $\theta(x)=0$. Note that $\widetilde{\mathbf P}_{\rm Geom}^{\rm C}$ in Eq. \eqref{eq:PGEOMC2} depends on the sign of $h$ at the order of $h^2$, which, as clarified later, is unimportant for the first-order PT.

Employing $\widetilde{\mathbf P}_{\rm Geom}^{\rm C}$ that locates at the outer side of $\partial V_{\rm res}$, Eq. \eqref{eq:MT1} is reformulated as
\begin{align}
\widetilde\Ev_{\rm p}(\rv;\omega)= \oint_{\partial V_{\rm res}} \greenOPF_{\rm u}(\rv,\rvb'+0^+\nv;\omega) \widetilde{\mathbf P}_{\rm Geom}^{\rm C}(\rvb';\omega) d^2\rvb',
\label{eq:MT3}
\end{align}
Inserting Eq. \eqref{eq:PGEOMC2} into Eq. \eqref{eq:MT3} and, then, applying the modal-expansion routines detailed in Sec. \ref{Sec:QNMEXP}, we find that ${\bf\mathcal{H}}_{\rm p}$ in Eq. \eqref{eq:H1storderb} is reformulated to ${\bf\mathcal{H}}_{\rm p}^{\rm C}$ with
\begin{align}
 {\bf\mathcal{H}}_{{\rm p}; nm}^{\rm C}=\underbrace{\mel{\left(\widetilde\Ev_{{\rm u};n}^{\rm bg}\right)^*}{
h \Delta\bm\varepsilon(\widetilde\omega_{{\rm u};m})}{\widetilde\Ev_{{\rm u};m}^{\rm res}}_{\scriptscriptstyle \partial V_{\rm res}}}_{\text{\normalsize$\propto h$}}
-
\underbrace{
\mel{\left(\widetilde\Ev_{{\rm u};n}^{\rm bg}\right)^*}{
\Delta \bm \varepsilon (\widetilde\omega_{{\rm u};m}) h \theta[-h] {\bf\mathcal{M}}_{\rm f-res}^{-1} {\bf \mathcal{K}} h }{\widetilde\Ev_{{\rm u};m}^{\rm bg}}_{\scriptscriptstyle \partial V_{\rm res}}}_{\text{\normalsize$\propto h^2$}}.
\label{eq:H1storderC}
\end{align}
Note that, for the diagonal components of ${\bf\mathcal{H}}_{\rm p}^{\rm C}$, the second term ($\propto h^2$) on the right-hand side of Eq. \eqref{eq:H1storderC} vanishes~\footnote{
Proof of $\mel{\left(\widetilde\Ev_{{\rm u};n}^{\rm bg}\right)^*}{
\Delta \bm \varepsilon (\widetilde\omega_{{\rm u};n}) h \theta[-h] {\bf\mathcal{M}}_{\rm f-res}^{-1} {\bf \mathcal{K}} h }{\widetilde\Ev_{{\rm u};n}^{\rm bg}}_{\scriptscriptstyle \partial V_{\rm res}}=0$:\\
\begin{align}
&\mel{\left(\widetilde\Ev_{{\rm u};n}^{\rm bg}\right)^*}{
\Delta \bm \varepsilon (\widetilde\omega_{{\rm u};n}) h \theta[-h] {\bf\mathcal{M}}_{\rm f-res}^{-1} {\bf \mathcal{K}}h }{\widetilde\Ev_{{\rm u};n}^{\rm bg}}_{\scriptscriptstyle \partial V_{\rm res}}\nonumber\\
&= \oint_{\partial\Omega} \left\{ \left[{\bf\mathcal{M}}_{\rm f-res}^{\rm T} (\widetilde\omega_{{\rm u};n}) \right]^{-1}\Delta\bm\varepsilon(\widetilde\omega_{{\rm u};n})h
\widetilde \Ev_{{\rm u};n}(\rv_{\scriptscriptstyle\partial V_{\rm res}}^{\rm bg})\right\}
\theta[-h]
\cdot
\left\{
 {\bf \mathcal{K}}h
\widetilde\Ev_{{\rm u};n}(\rv_{\scriptscriptstyle\partial V_{\rm res}}^{\rm bg})
\right\} d^2\rvb \nonumber\\
& = -\oint_{\partial V_{\rm res}} \underbrace{\left\{ \left[{\bf\mathcal{M}}_{\rm f-res}^{\rm T} (\widetilde\omega_{{\rm u};n}) \right]^{-1}\Delta\bm\varepsilon(\widetilde\omega_{{\rm u};n})h
\widetilde \Ev_{{\rm u};n}(\rv_{\scriptscriptstyle\partial V_{\rm res}})\cdot{(\Iv-\nv\nv)}\right\}}_{\text{\normalsize$\equiv \mathbf A$}}
\theta[-h(\rvb)]
\cdot
\left\{
\bm\nabla_{\scriptscriptstyle\parallel}
\underbrace{\left[{\bf\mathcal{M}}_{\rm f-res}^{\rm T} (\widetilde\omega_{{\rm u};n}) \right]^{-1}\Delta\bm\varepsilon(\widetilde\omega_{{\rm u};n})h
\widetilde\Ev_{{\rm u};n}(\rv_{\scriptscriptstyle\partial V_{\rm res}}^{\rm bg})
\cdot \nv}_{\text{\normalsize$\equiv \mathbf B$}}
\right\} d^2\rvb\nonumber\\
& -\oint_{\partial V_{\rm res}} \underbrace{\left\{ \left[{\bf\mathcal{M}}_{\rm f-res}^{\rm T} (\widetilde\omega_{{\rm u};n}) \right]^{-1}\Delta\bm\varepsilon(\widetilde\omega_{{\rm u};n})h
\widetilde \Ev_{{\rm u};n}(\rv_{\scriptscriptstyle\partial V_{\rm res}})\cdot\nv\right\}}_{\text{\normalsize$\equiv \mathbf B$}}
\theta[-h]
\left\{
\bm\nabla_{\scriptscriptstyle\parallel}\cdot
\underbrace{\left(\left[{\bf\mathcal{M}}_{\rm f-res}^{\rm T} (\widetilde\omega_{{\rm u};n}) \right]^{-1}\Delta\bm\varepsilon(\widetilde\omega_{{\rm u};n})h
\widetilde\Ev_{{\rm u};n}(\rv_{\scriptscriptstyle\partial V_{\rm res}}^{\rm bg})
\cdot (\Iv-\nv\nv) \right)}_{\text{\normalsize$\equiv \mathbf A$}}
\right\} d^2\rvb \nonumber\\
&= -\oint_{\partial V_{\rm res}} \bm\nabla_{\scriptscriptstyle\parallel} \cdot \left[\mathbf A  B \theta(-h)\right]
 d^2\rvb\nonumber\\
&=0.\nonumber
\end{align}
}.

At this stage, we thus have proved the equivalence between Eq. \eqref{eq:H1storderb} and  Eq. \eqref{eq:H1storderC}. Further, we note that the first-order PT is only obliged to be accurate to the first order of $h$. This thus implies that, dropping the second-order term ($\propto h^2$) on the right-hand side of Eq. \eqref{eq:H1storderC}, the theory still retains the first-order accuracy. Therefore, within the framework of the first-order PT, it is justified to approximate ${\bf\mathcal{H}}_{{\rm p}; nm}^{\rm C}$ with Eq. \eqref{eq:H1storder}~\footnote{Similarly, by moving $\widetilde {\mathbf P}_{\rm Geom}$ from the background side of $\partial V_{\rm res}$ to the resonator side, we could have another equivalent expression,
\begin{align}
\mathcal{H}_{{\rm p}; nm}\simeq
\mel{\left(\widetilde\Ev_{{\rm u};n}^{\rm res}\right)^*}{
h \Delta\bm\varepsilon(\widetilde\omega_{{\rm u};m})}{\widetilde\Ev_{{\rm u};m}^{\rm bg}}_{\scriptscriptstyle \partial V_{\rm res}}.
\nonumber
\end{align}}, i.e.,
\begin{align}
\mathcal{H}_{{\rm p}; nm}\simeq
\mel{\left(\widetilde\Ev_{{\rm u};n}^{\rm bg}\right)^*}{
h \Delta\bm\varepsilon(\widetilde\omega_{m})}{\widetilde\Ev_{{\rm u};n}^{\rm res}}_{\scriptscriptstyle \partial V_{\rm res}}.
\nonumber
\end{align}

%In comparison with Eq. \eqref{eq:H1storderb}, one obvious advantage of using Eq. \eqref{eq:H1storder} or Eq. \eqref{eq:H1storderc} is that the latter two equations are analytic with respect to $h$, ensuring that $\left.\delta\widetilde\omega_{\rm p}/\delta h\right|_{h=0}$ and $\left.\delta \ket{\bm\alpha}/\delta h\right|_{h=0}$ are well mathematically defined.

\section{First-order Perturbation Theory: Numerical Results}

\begin{figure}[!htp]
\centering
\includegraphics[width=10cm]{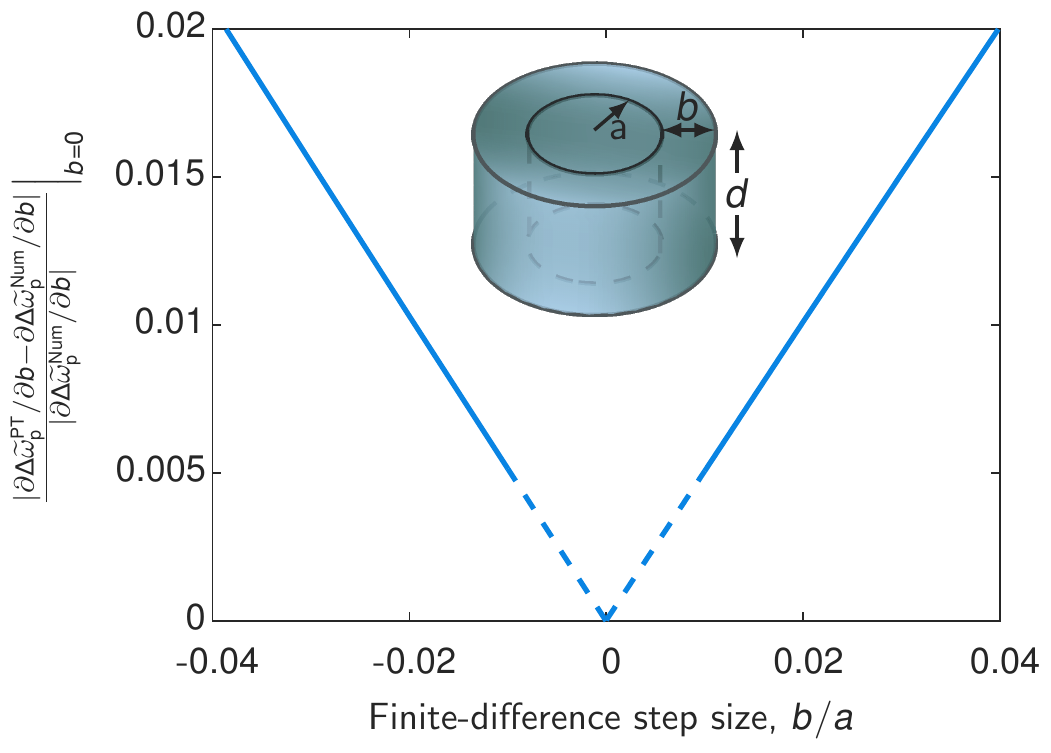}
\caption{
{\bf Numerical validation of the first-order PT formula Eq. (3)} for a silicon nanorod in air. Solid lines represent the computed difference between theoretical and numerical predictions of first-order derivatives of perturbed-QNM frequencies with respect to perturbation parameter.
Dashed lines represent linear extrapolations of the computed values as finite-difference step size goes to zero.
The nanorod, with radius $a=200\,\rm nm$, height $d=400\,\rm nm$ and permittivity $\varepsilon_{\rm si}=12.96$, is deformed by a uniform radial change $h=b$ (see the inset). A (1,1,1) QNM, with $\widetilde\omega_{\rm u}=0.74-0.034{\rm i}\,\rm eV$, is considered, c.f. Fig. 2(a) (main text) for its modal profiles.
}
\label{Fig:1stAgTikz}
\end{figure}

{\bf Example 1---}First-order PTs become exact in the limit of infinitely small perturbations, therein implying that they predict the slope of perturbation-induced change. Therefore, the most straightforward way to validate first-order perturbation formula is to compute the first-order derivative of perturbed quantity with respect to perturbation parameter, and to compare the numerical results with the theoretical predictions. Following this methodology, we revisit the numerical example studied in Fig. 2 in the main text: A silicon nanorod is deformed by a uniform radial change $h=b$. Theoretically, Eq. (3) gives $\left.{\partial \Delta \widetilde\omega}/{\partial b}\right|_{b=0}=-\widetilde\omega_{{\rm u};n}\mel{\left(\widetilde\Ev_{{\rm u};n}^{\rm bg}\right)^*}{
\Delta\bm\varepsilon(\widetilde\omega_{n})}{\widetilde\Ev_{{\rm u};n}^{\rm res}}_{\scriptscriptstyle \partial V_{\rm res}}$. Numerically, we use the QNMEig solver to compute perturbed modes, and then the first-order derivatives are evaluated with the finite-difference approach
\begin{align}
\left.\frac{\partial \Delta \widetilde\omega}{\partial b}\right|_{b=0}\simeq \frac{\widetilde\omega_{\rm p}(b)-\widetilde\omega_{\rm u}}{b}.
\label{eq:1st-D}
\end{align}
We compute the relative difference between $\left.{\partial \Delta \widetilde\omega^{\rm PT}}/{\partial b}\right|_{b=0}$ and $\left.{\partial \Delta \widetilde\omega^{\rm Num}}/{\partial b}\right|_{b=0}$---with superscripts “PT" and “Num" indicating theoretical and numerical predictions, respectively---as a function of $b$. The results plotted in Figure \ref{Fig:1stAgTikz} show that, as $b$ goes to zero, the numerical values of the first-order derivatives using Eq. \eqref{eq:1st-D} become more accurate, and, accordingly, the difference between the theoretical and numerical results approaches zero. This conclusively validates the soundness of the first-order PT.

\begin{figure}[!htp]
\centering
\includegraphics[width=16cm]{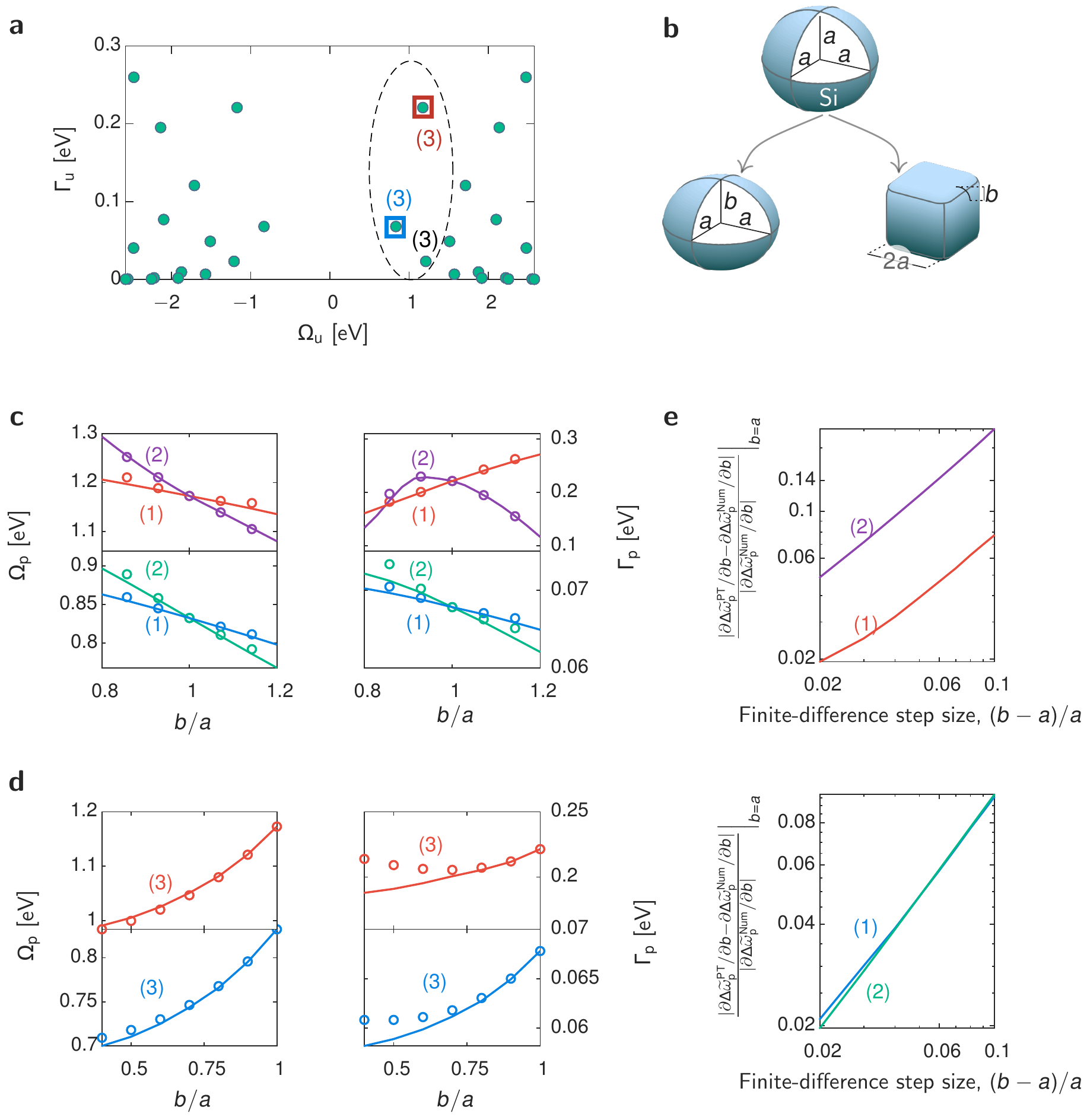}
\caption{
{\bf Validation of the first-order PT of Eq. (2)} for a silicon nanosphere in air. The nanosphere has a diameter of 400 nm and a permittivity of $\varepsilon_{\rm si}=12.96$.
{\bf a.} QNM eigenfrequencies. The 9 QNMs used for validating the first-order PT are surrounded by an ellipse and the degeneracy factors are given in parenthesis
{\bf b.} Sketch of deforming nanosphere into spheroids and cuboids.
{\bf c-\bf d.} Eigenfrequencies, $\widetilde\omega_{\rm p}\equiv\Omega_{\rm p}-i\Gamma_{\rm p}/2$, of perturbed QNMs of spheroids ({\bf c}) and cuboids ({\bf d}) as a function of the aspect ratio $b/a$, contrasting the theoretical (solid lines) and numerical (circles) predictions. Eigenfrequencies of unperturbed QNMs are marked with squares in {\bf a}. Tiny deformations correspond to $a\simeq b$.
{\bf e.} Difference of first-order derivatives between theoretical and numerical predictions of QNM frequencies for QNMs of spheroids illustrated in {\bf c}.
}
\label{Fig:1stSiTikz}
\end{figure}

{\bf Example 2---}We continue to test the first-order PT by considering a silicon nanosphere (400-nm diameter) in air that is deformed into spheroids and cuboids. The QNM frequencies of the spheroids and cuboids are computed with Eqs. (2), retaining only 9 QNMs whose frequencies are plotted in Fig. \ref{Fig:1stSiTikz}(a) (inside the dashed ellipse). Tracking the perturbation-induced frequency shifts of those selected QNMs [marked with squares in Fig. \ref{Fig:1stSiTikz}(a)], Fig. \ref{Fig:1stSiTikz}(c) and (d) plot the modal frequencies as functions of aspect ratio $b/a$ [$a=200\,\rm nm$; see Fig. \ref{Fig:1stSiTikz}(b)] for the spheroids and cuboids, respectively. The results show good agreements between the theoretical and numerical predictions, wherein the deformations result in maximal volume changes over 30\%$\sim$50\%.
We further examine the QNM frequencies of the spheroids in Figs. \ref{Fig:1stSiTikz}(c), and compute the first-order derivatives of the frequency shifts with respect to the deformation parameter $b$ at the non-perturbation point $b=a$. Note that, in this case, since the theoretical predictions of the perturbed frequencies are obtained by solving the eigensolutions of a $9 \times 9$ matrix, the associated first-order derivatives are evaluated with the finite-difference approach similarly as the numerical predictions using the QNMEig solver. The results plotted in
Fig. \ref{Fig:1stSiTikz}(e) show that the differences between the theoretical and numerical values of the first-order derivatives approach zero as the finite-difference step size approaches zero.

\section{Second-order Perturbation Theory}
\label{Sec:2ndQPT_1}

%{The results in this section do not relate directly with the main text that focuses on the first-order QNM PT. We here include these results to demonstrate that the proposed QNM PT can be used for high-order perturbative analysis with improved accuracy, while at the cost of requiring more QNMs.}

\subsection{Theoretical Analysis}
\label{Sec:2dnQPT}

By retaining both $\propto h$ and $\propto h^2$ terms in $\widetilde {\mathbf P}_{\rm Geom}$ [Eq. \eqref{eq:PGEOM}], the first-order PT can be generalized to the second-order one, wherein ${\bf\mathcal {H}}_{\rm p}$ is given by
\begin{align}
{\bf\mathcal {H}}_{{\rm p};nm} & \simeq \underbrace{\mel{\left(\widetilde\Ev_{{\rm u};n}^{-\sigma[h]}\right)^*}{
h \Delta\bm\varepsilon(\widetilde\omega_{{\rm u};m})}{\widetilde\Ev_{{\rm u};m}^{\sigma[h]}}_{\scriptscriptstyle \partial V_{\rm res}}}_{\text{\normalsize $\propto h$}}+
\underbrace{\mel{\left(\widetilde\Ev_{{\rm u};n}^{-\sigma[h]}\right)^*}{
h^2 \kappa_{\rm m}\Delta\bm\varepsilon(\widetilde\omega_{{\rm u};m})}{\widetilde\Ev_{{\rm u};m}^{\sigma[h]}}_{\scriptscriptstyle \partial V_{\rm res}}}_{\text{\normalsize $\propto h^2$}}\nonumber\\
&+
\underbrace{\frac{1}{2}\mel{\left(\widetilde\Ev_{{\rm u};n}^{-\sigma[h]}\right)^*}{
h^2 \Delta\bm\varepsilon(\widetilde\omega_{{\rm u};m})}{\nv\cdot\bm\nabla\widetilde\Ev_{{\rm u};m}^{\sigma[h]}}_{\scriptscriptstyle \partial V_{\rm res}}
+\frac{1}{2}\mel{\nv\cdot\bm\nabla\left(\widetilde\Ev_{{\rm u};n}^{-\sigma[h]}\right)^*}{
h^2 \Delta\bm\varepsilon(\widetilde\omega_{{\rm u};m})}{\widetilde\Ev_{{\rm u};m}^{\sigma[h]}}_{\scriptscriptstyle \partial V_{\rm res}}}_{\text{\normalsize $\propto h^2$}}.
\label{eq:H2ndorder}
\end{align}
Similar to first-order PTs, any correct second-order PTs should guaranty that perturbed quantities are analytic with respect to perturbation parameter. By relating fields at two sides of $\partial V_{\rm res}$, we have proven this property for the first-order PT in Sec. \ref{Sec:1stQP_Theory}. Generalizing the proof for the second-order PT, however, requires more tedious algebra manipulations. We here refrain from digging into this problem comprehensively. Nevertheless, we shall provide the following evidences to validate the second-order PT: (1) a proof that the second-order frequency shift for a single mode is analytic with respect to perturbation parameter (at the end of this subsection); (2) numerical results showing that the second-order PT improves the prediction accuracy for perturbed modal frequencies, see Sec. \ref{Sec:2dnQPTNR}.

\subsubsection*{\bf Proof of analyticity of second-order frequency shift}

Consider an unperturbed mode with eigenfrequency $\widetilde\omega_{{\rm u};n}$. We derive the perturbation-induced frequency shift in the limit of vanishing perturbation to the order of $h^2$:
\begin{align}
\Delta\widetilde\omega_{n} & \simeq   -\widetilde\omega_{{\rm u};n} \left[ {{\bf\mathcal {H}}_{{\rm p};nn}}- {\left( {\bf\mathcal {H}}_{{\rm p};nn} \right)^2-\sum_{m \ne n} \frac{\widetilde\omega_{n}}{\widetilde\omega_{{\rm u};n}-\widetilde\omega_m} {\bf\mathcal {H}}_{{\rm p};nm}  {\bf\mathcal {H}}_{{\rm p};mn}}\right] \nonumber\\
& \simeq
-\widetilde\omega_{{\rm u};n}\left[
\underbrace{\mel{\left(\widetilde\Ev_{{\rm u};n}^{\rm bg}\right)^*}{
h \Delta\bm\varepsilon(\widetilde\omega_{n})}{\widetilde\Ev_{{\rm u};n}^{\rm res}}_{\scriptscriptstyle \partial V_{\rm res}}}_{\text{\normalsize $\propto h$}}
\underbrace{-\mel{\left(\widetilde\Ev_{{\rm u};n}^{\rm bg}\right)^*}{
h \Delta\bm\varepsilon(\widetilde\omega_{{\rm u};n})}{\widetilde\Ev_{{\rm u};n}^{\rm res}}_{\scriptscriptstyle \partial V_{\rm res}}^2}_{\text{\normalsize $\propto h^2$}}\right.\nonumber\\
&\quad\underbrace{\mel{\left(\widetilde\Ev_{{\rm u};n}^{\rm bg}\right)^*}{
h^2 \kappa_{\rm m}\Delta\bm\varepsilon(\widetilde\omega_{{\rm u};n})}{\widetilde\Ev_{{\rm u};n}^{\rm res}}_{\scriptscriptstyle \partial V_{\rm res}}
+\frac{1}{2}\mel{\left(\widetilde\Ev_{{\rm u};n}^{\rm bg}\right)^*}{
h^2 \Delta\bm\varepsilon(\widetilde\omega_{{\rm u};n})}{\nv\cdot\bm\nabla\widetilde\Ev_{{\rm u};n}^{\rm res}}_{\scriptscriptstyle \partial V_{\rm res}}
+\frac{1}{2}\mel{\nv\cdot\bm\nabla\left(\widetilde\Ev_{{\rm u};n}^{\rm bg}\right)^*}{
h^2 \Delta\bm\varepsilon(\widetilde\omega_{{\rm u};n})}{\widetilde\Ev_{{\rm u};n}^{\rm res}}_{\scriptscriptstyle \partial V_{\rm res}}}_{\text{\normalsize $\propto h^2$}}\nonumber\\
&\left. \quad\underbrace{-\sum_{m \ne n} \frac{\widetilde\omega_{{\rm u};n}}{\widetilde\omega_{{\rm u};n}-\widetilde\omega_{{\rm u};m}}
\mel{\left(\widetilde\Ev_{{\rm u};n}^{\rm bg}\right)^*}{
h \Delta\bm\varepsilon(\widetilde\omega_{{\rm u};m})}{\widetilde\Ev_{{\rm u};m}^{\rm res}}_{\scriptscriptstyle \partial V_{\rm res}}
\mel{\left(\widetilde\Ev_{{\rm u};m}^{\rm bg}\right)^*}{
h \Delta\bm\varepsilon(\widetilde\omega_{{\rm u};n})}{\widetilde\Ev_{{\rm u};n}^{\rm res}}_{\scriptscriptstyle \partial\Omega_{{\rm u}}}}_{\text{\normalsize $\propto h^2$}}\right].
\label{eq:omega2nd}
\end{align}
Here, in the first line, we take ${{\bf\mathcal {H}}_{{\rm p}}}$ from Eq. \eqref{eq:H2ndorder}; in the second line, we use ${\bf\mathcal {H}}_{{\rm p}}^{(C)}$ in Eq. \eqref{eq:H1storderC} to replace the term
${\mel{\left(\widetilde\Ev_{{\rm u};n}^{-\sigma[h]}\right)^*}{
h \Delta\bm\varepsilon(\widetilde\omega_{{\rm u};m})}{\widetilde\Ev_{{\rm u};m}^{\sigma[h]}}_{\scriptscriptstyle \partial V_{\rm res}}}$ \footnote{Note that the term $\propto h^2$ in ${\bf\mathcal {H}}_{{\rm p}; nn}^{(C)}$ vanishes, see footnote 7 for derivations.}, and only keep the terms $\propto h$ and $\propto h^2$.
Examining Eq. \eqref{eq:omega2nd}, it is obvious that $\Delta\widetilde\omega_{n}$ is analytic with respect to $h$.
%\begin{align}
%\left.\frac{\delta {\widetilde\omega_{{\rm p}; n}}[h]/\widetilde \omega_ {{\rm u};n}}{\delta h}\right|_{h=0} & \simeq \Ev_{{\rm u};n}(\rvb^{\rm bg})\Delta\bm\varepsilon(\widetilde\omega_{{\rm u};n})\Ev_{{\rm u};n}(\rvb^{\rm res})\nonumber\\
%\left.\frac{\delta^2 {\widetilde\omega_{{\rm p}; n}}[h]/\widetilde \omega_{{\rm u};n}}{\delta h^2}\right|_{h=0} & =
%\left[\Ev_{{\rm u};n}(\rvb^{\rm bg})\Delta\bm\varepsilon(\widetilde\omega_{{\rm u};n})\Ev_{{\rm u};n}(\rvb^{\rm res})\right]^2
%-\left[2\Ev_{n}(\rvb^{\rm bg})\kappa_{\rm m}\Delta\bm\varepsilon(\widetilde\omega_{{\rm u};n})\Ev_{{\rm u};n}^{\rm res}(\rvb)+\right.\nonumber\\
%&\quad\left.
%         \left(\nv\cdot\nablav\Ev_{{\rm u};n}(\rvb^{\rm bg})\right)\Delta\bm\varepsilon(\widetilde\omega_{{\rm u};n})\Ev_{{\rm u};n}(\rvb^{\rm res})+
%         \Ev_{{\rm u};n}(\rvb^{\rm bg})\Delta\bm\varepsilon(\widetilde\omega_{{\rm u};n})
%         \left(\nv\cdot\nablav\Ev_{{\rm u};n}(\rvb^{\rm res})\right)
%         \right]{ \delta ^2(\rv-\rvb)}\nonumber\\
%&+\sum_{m\ne n}
%\frac{2\widetilde\omega_{{\rm u};n}}{\widetilde\omega_{{\rm u};n}-\widetilde\omega_{{\rm u};m}}
%\Ev_{{\rm u};n}(\rvb^{\rm res})\Delta\bm\varepsilon(\widetilde\omega_{{\rm u};m})\Ev_{{\rm u};m}(\rvb^{\rm bg})
%\Ev_{{\rm u};m}(\rvb^{\rm bg})\Delta\bm\varepsilon(\widetilde\omega_{{\rm u};n}) \Ev_{{\rm u};n}(\rvb^{\rm res})\nonumber,
%\end{align}
%where $\delta ^2(\rv-\rvb)$ denotes a two-dimensional Dirac delta function defined at $\partial V_{\rm res}$.

\subsection{Numerical Results}
\label{Sec:2dnQPTNR}

In Fig. 2 in the main text, we have validated the second-order PT in the limit of vanishing perturbation. In this subsection, we provide extra numerical results and test the second-order PT.

\begin{figure}[!t]
\centering
\includegraphics[width=15cm]{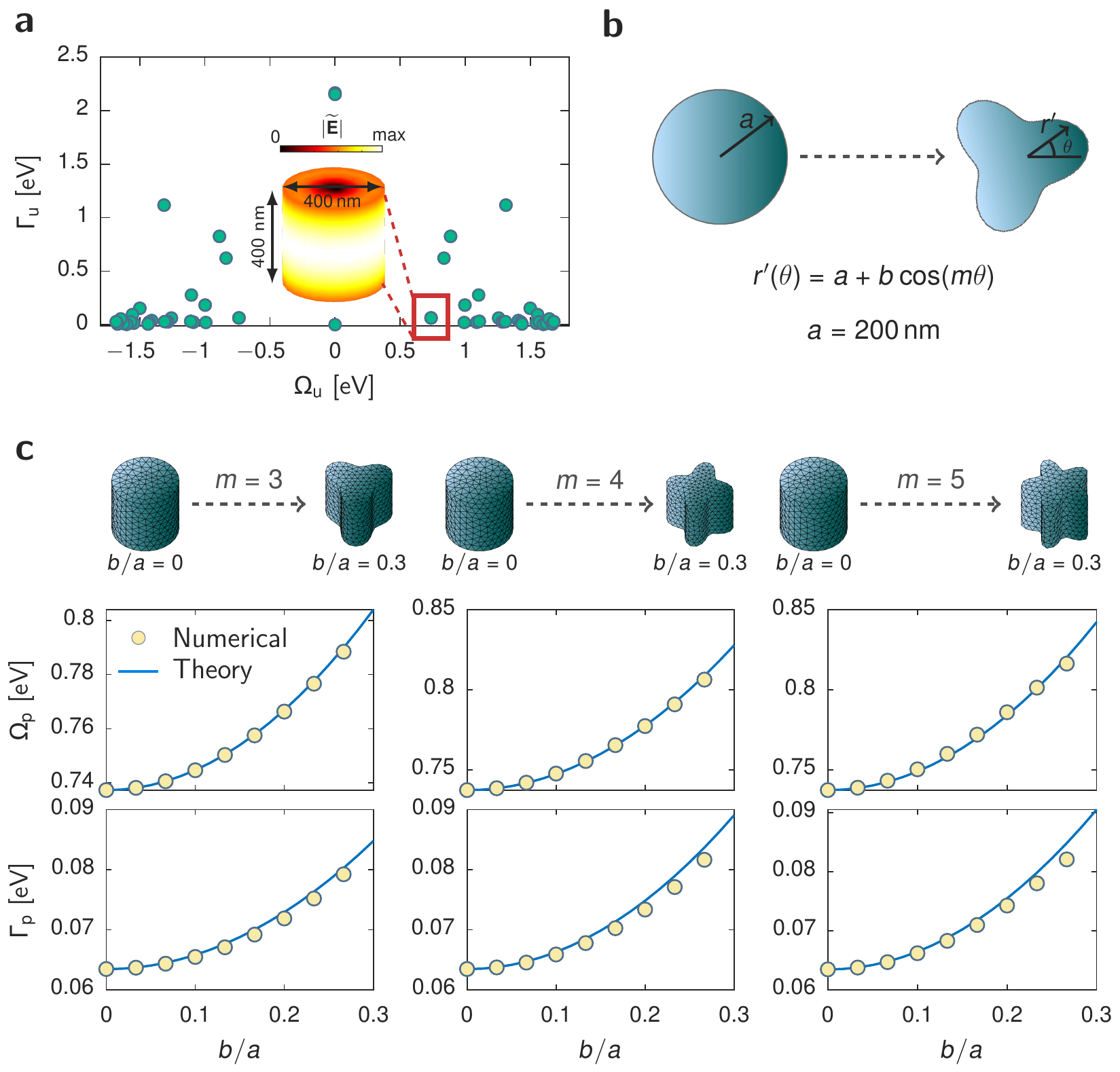}
\caption{
{\bf Validation of the second-order PT of Eq. \eqref{eq:omega2nd}} for a silicon rod in air. The rod has a diameter of 400 nm, a height of 400 nm and a relative permittivity of 12.96.
{\bf a.} Eigenfrequency spectrum of modes supported by the rod.
The inset depicts amplitude of electric field of a selected QNM on the (air side) nanorod surface.
{\bf b.} Sketch of studied geometrical deformations.
{\bf c.} Modal frequency of perturbed QNMs as functions of deformation parameter $b/a$ for a class of deformations with $m=3,4,5$ (c.f. {\bf b}).
}
\label{Fig:2ndSiRod}
\end{figure}

{\bf Example 1---}Consider the same silicon nanorod as in Fig. 2. The eigenfrequency spectrum of modes is plotted in Fig. \ref{Fig:2ndSiRod}(a). We deform the nanorod by changing its radius $a= 200\, \rm nm$ to
\begin{align}
r'(\theta)=a+b\cos(m\theta), \nonumber
\end{align}
where $m$ is an integer and $\theta$ denotes the azimuthal angle in the cylindrical coordinate system, see Fig. \ref{Fig:2ndSiRod}(b). We select a QNM marked with square in Fig. \ref{Fig:2ndSiRod}(a) for the following numerical study. A group of deformations with $m=3,4,5$ is considered, in which the modal symmetry results in the vanishing of the first-order correction, and, accordingly, the second-order correction becomes dominant. We compute the eigenfrequencies of the perturbed QNM as functions of $b$ with Eq. \eqref{eq:omega2nd}. Note that the modal truncation in the series summation of Eq. \eqref{eq:omega2nd} inevitably induces numerical error. To mitigate the truncation error, we exploit the sum rule for non-dispersive materials~\cite{Muljarov:2016}
\begin{align}
\sum_{n}^{\widetilde\omega_n \ne 0}
\frac{\widetilde\Ev_n\otimes\widetilde\Ev_n}{\widetilde\omega_{{\rm u};n}}=0.
\label{eq:sum_rule}
\end{align}
Using Eq. \eqref{eq:sum_rule} and defining ${\bf\mathcal {H}}_{{\rm p};nm}^{(1)}\equiv \mel{\left(\widetilde\Ev_{{\rm u};n}^{\rm bg}\right)^*}{
h \Delta\bm\varepsilon}{\widetilde\Ev_{{\rm u};m}^{\rm res}}_{\scriptscriptstyle \partial V_{\rm res}}$, the series summation in Eq. \eqref{eq:omega2nd} can be reformulated as
\begin{align}
&\sum_{m \ne n} \frac{\widetilde\omega_{{\rm u};n}}{\widetilde\omega_{{\rm u};n}-\widetilde\omega_{{\rm u};m}} {\bf\mathcal {H}}_{{\rm p};nm}^{(1)}  {\bf\mathcal {H}}_{{\rm p};mn}^{(1)}\nonumber\\
&= \sum_{m \ne n}^{\widetilde\omega_{{\rm u};m}\ne 0} \frac{\widetilde\omega_{{\rm u};n}}{\widetilde\omega_{{\rm u};n}-\widetilde\omega_{{\rm u};m}} {\bf\mathcal {H}}_{{\rm p};nm}^{(1)}  {\bf\mathcal {H}}_{{\rm p};mn}^{(1)}+
   {\sum_{m}^{\widetilde\omega_{{\rm u};m}= 0} {\bf\mathcal {H}}_{{\rm p};nm}^{(1)}  {\bf\mathcal {H}}_{{\rm p};mn}^{(1)}}\nonumber\\
&= \sum_{m \ne n}^{\widetilde\omega_{{\rm u};m}\ne 0} \frac{\widetilde\omega_{{\rm u};n}}{\widetilde\omega_{{\rm u};n}-\widetilde\omega_{{\rm u};m}} {\bf\mathcal {H}}_{{\rm p};nm}^{(1)}  {\bf\mathcal {H}}_{{\rm p};mn}^{(1)}+
   \underbrace{\sum_{m}^{\widetilde\omega_{{\rm u};m}\ne 0}  \frac{\widetilde\omega_{{\rm u};n}}{\widetilde\omega_{{\rm u};m}} {\bf\mathcal {H}}_{{\rm p};nm}^{(1)}  {\bf\mathcal {H}}_{{\rm p};mn}^{(1)}}_{\text{\normalsize application of the sum rule}}+
   {\sum_{m}^{\widetilde\omega_{{\rm u};m}= 0} {\bf\mathcal {H}}_{{\rm p};nm}^{(1)}  {\bf\mathcal {H}}_{{\rm p};mn}^{(1)}}  \nonumber\\
   &= \underbrace{\sum_{m \ne n}^{\widetilde\omega_{{\rm u};m}\ne 0} \frac{\widetilde\omega_{{\rm u};n}^2}{(\widetilde\omega_{{\rm u};n}-\widetilde\omega_{{\rm u};m})\widetilde\omega_{{\rm u};m}} {\bf\mathcal {H}}_{{\rm p};nm}^{(1)}  {\bf\mathcal {H}}_{{\rm p};mn}^{(1)}}_
   {\text{\normalsize contribution from transverse modes}}+
 \underbrace{\sum_{m}^{\widetilde\omega_{{\rm u};m}= 0} {\bf\mathcal {H}}_{{\rm p};nm}^{(1)}  {\bf\mathcal {H}}_{{\rm p};mn}^{(1)}}_{\text{\normalsize contribution from zero-frequency modes}}+
   \left({\bf\mathcal {H}}_{{\rm p};nn}^{(1)}\right)^2.
\label{eq:SeriesE}
\end{align}

Figure \ref{Fig:2ndSiRod}(c) plots the computed modal frequencies of the perturbed modes. The summation rule, summarized in Eq. \eqref{eq:SeriesE}, is used, retaining 90 transverse modes plotted in Fig. \ref{Fig:2ndSiRod}(a), and the contribution from zero-frequency modes is taken into account with the technique detailed in Sec. \ref{Sec:TRemarksLM}. We observe that the theoretical and numerical results agree well with each other even.

\begin{figure}[!htp]
\centering
\includegraphics[scale=1]{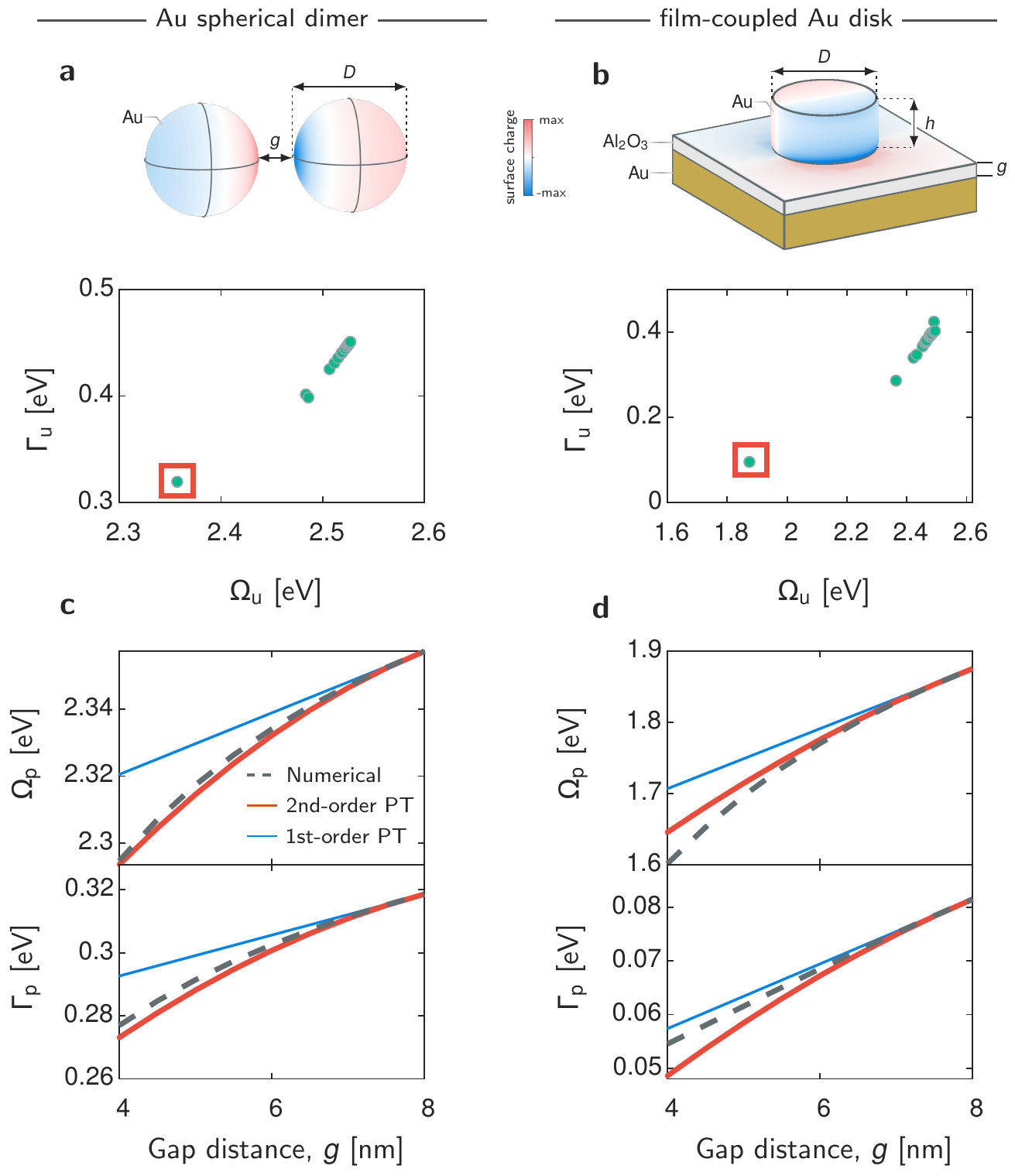}
\caption{
{\bf Validation of the second-order PT of Eq. \eqref{eq:omega2nd}} for gap-plasmon structures. {\bf a-b} Modal profiles (top panels) and QNM frequencies (bottom panels) for two representative gap plasmonic structures: {(\bf a)} Au sphere dimer with diameter $D = 60\,\rm nm$; ({\bf b}) Au nanodisk
with diameter $D = 60\,\rm nm$ and height $h=60\,\rm nm$ on an Au substrate covered by a thin dielectric film. The gap distances $g$ for both structures are 8 nm. The dielectric layer in {\bf b} has a relative permittivity of $2.25$.
{\bf c-d} Modal eigenfrequencies, $\widetilde\omega_{\rm p}\equiv\widetilde\Omega_{\rm p}-{\rm i}\Gamma_{\rm p}/2$, of perturbed dipole modes as functions of gap distance, contrasting the first and second-order PT results, and the numerical results obtained with the QNMEig solver. The second-order PT retains $40$ and $52$ transverse QNMsfor the dimer and nanodisk structures, respectively.
%{, where the factor $2$ accounts for modal complex-conjugated-counterparts $\left\{-\widetilde\omega_{\rm u}^*,\widetilde\Ev_{\rm u}^*\right\}$}.
Moreover, the contribution from zero-frequency modes is included.
}
\label{Fig:2ndAuDisk}
\end{figure}

{\bf Example 2---}We study two representative gap-plasmon structures: (1) dimer of Au nanospheres in air and (2)
Au nanodisk on a Au substrate covered with a thin layer dielectric medium, see the top panels in Figs. \ref{Fig:2ndAuDisk} (a) and (b), respectively. The Au permittivity is approximated by the Lorentz-Drude model, which is the same as in Fig. 4 in the main text. Figures \ref{Fig:2ndAuDisk}(a)-(b) (bottom panels) plot the eigenfrequency spectra of modes of the two structures with gap distance $g=8\,\rm nm$.
We choose the fundamental dipole modes for the numerical studies, whose frequencies are marked with red squares. The rotational symmetry is exploited for modal computations. Specifically, we consider the modes with rotational invariance and azimuthal dependence $\exp(i\phi)$ for the dimer and disk structures, respectively.

We employ both the first- and second-order PTs to compute the eigenfrequencies of the perturbed dipole plasmonic modes as the gap distance decreases. Note that, since Au is dispersive, the sum rule is not used. Moreover, the second-order PT includes 40 and 52 transverse modes for the dimer and disk structures, respectively. The contribution from zero-frequency modes is also included with the technique detailed in Sec. \ref{Sec:TRemarksLM}.
The results plotted in Figs. \ref{Fig:2ndAuDisk}(c)-(d) show that, the first-order PT gives accurate results when $g$ is close to 8 nm (unperturbed value); with the second-order PT, the numerical accuracy is further improved.

%The Au permittivity is approximated by the Lorentz-Drude model:
%\begin{align}
%\varepsilon_{\rm Au}(\omega)=\varepsilon_{\infty}\left[1-\frac{\omega_{\rm p; Drude}^2}{\omega^2+i\gamma_{\rm Drude}}-\frac{\omega_{\rm p; Lorentz}^2}{\omega^2-\omega_0^2+i\gamma_{\rm Lorentz}}\right]\nonumber
%\end{align}
%where $\varepsilon_{\infty}=6$, $\hbar\omega_{\rm p; Drude}=3.54\,[\rm eV]$, $\hbar\omega_{\rm p; Lorentz}=1.50\,[\rm eV]$, $\hbar \gamma_{\rm Drude}=0.04\,[\rm eV]$, $\hbar \gamma_{\rm Lorentz}=0.88\,[\rm eV]$ and $\hbar \omega_{0}=3.01\,[\rm eV]$.

\subsection{Technical Remarks on Including Contributions from Zero-frequency Modes}
\label{Sec:TRemarksLM}

The second-order frequency shift formula involves a series summation that includes zero-frequency modes with $\nablav\times\widetilde{\Ev}_{{\rm u};n}=0$ and $\widetilde\omega_{{\rm u}; n}=0$. Referring to Eq. \eqref{eq:omega2nd}, the frequency shift contributed from these zero-frequency mode, denoted by $\Delta\widetilde\omega_{n}^{\rm L}$, has the expression
\begin{align}
\Delta\widetilde\omega_{n}^{\rm L}
&{=}\widetilde\omega_{{\rm u};n} {\sum_{m}^{\widetilde\omega_{{\rm u};m}= 0}
\mel{\left(\widetilde\Ev_{{\rm u};n}^{\rm bg}\right)^*}{
h \Delta\bm\varepsilon(0)}{\widetilde\Ev_{{\rm u};m}^{\rm res}}_{\scriptscriptstyle \partial V_{\rm res}}
\mel{\left(\widetilde\Ev_{{\rm u};m}^{\rm bg}\right)^*}{
h \Delta\bm\varepsilon(\widetilde\omega_{{\rm u};n})}{\widetilde\Ev_{{\rm u};n}^{\rm res}}_{\scriptscriptstyle \partial V_{\rm res}} } \nonumber\\
&{=}\widetilde\omega_{{\rm u};n}\mel{\left(\widetilde\Ev_{{\rm u};n}^{\rm bg}\right)^*}{
h \Delta\bm\varepsilon(0)}{\widetilde\Ev_{{\rm static};n}^{\rm res}}_{\scriptscriptstyle \partial V_{\rm res}},
\label{eq:EVL1}
\end{align}
where
\begin{align}
\ket{\Ev_{{\rm static}; n}^{\rm res}}\equiv {\sum_{m}^{\widetilde\omega_{{\rm u};m}= 0}\ket{\Ev_{{\rm u}; m}^{\rm res}}\mel{\left(\widetilde\Ev_{{\rm u};m}^{\rm bg}\right)^*}{
h \Delta\bm\varepsilon(\widetilde\omega_{{\rm u};n})}{\widetilde\Ev_{{\rm u};n}^{\rm res}}_{\scriptscriptstyle \partial V_{\rm res}} } .
\label{eq:EVL2}
\end{align}

Equation \eqref{eq:EVL1} suggests that the contribution from the zero-frequency modes can be taken into account by computing the newly defined fields $\widetilde\Ev_{{\rm static};n}^{\rm res}$. Further, by applying the modal expansion of $\greenOPF_{\rm u}$, Eq. \eqref{eq:GQNMEXP}, $\widetilde\Ev_{{\rm static};n}^{\rm res}$ can be reformulated in a different form that is convenient for numerical implementation,
\begin{align}
\Ev^{\rm res}_{{\rm static}; n}=\lim_{\omega\to 0} \omega^2 {\oint_{\partial V_{\rm res}} \greenOPF_{\rm u}(\rvb+0^-\nv, {\rvb'+0^+\nv})h \Delta\bm\varepsilon(\widetilde\omega_{{\rm u};n})
\widetilde\Ev_{{\rm u};n}({\rvb'+0^-\nv}) d^2\rvb'}_{}.
\end{align}
\begin{itemize}
\item
Apparently, $\Ev^{\rm res}_{{\rm static}; n}$ represents electric fields at the inner (resonator) side of $\partial V_{\rm res}$ induced by the surface polarization---$h\Delta\varepsilon(\widetilde\omega_{{\rm u};n})\widetilde\Ev_{{\rm u}; n}(\rvb+0^-\nv)$ that locates at the outer (background) side of $\partial V_{\rm res}$---at zero frequency
\end{itemize}
Based on the above interpretation, we compute $\widetilde\Ev_{{\rm static};n}^{\rm res}$ by simply solving an electrostatic problem with the boundary element method.

\section{Reconstruction of Optical Responses}
\label{sec:ReConTheory}

Reconstructing optical responses of perturbed nanoresonators---driven by external stimuli---with unperturbed modes is particularly useful for inverse design, wherein optimized objectives are often given in terms of spectral responses. By generalizing the PT formalism that predicts perturbed eigensolutions, we give Eq. (4) in the main text to predict optical responses, and numerically evidence its accuracy in Fig. 3(c). In this section, we shall present the derivation details of Eq. (4) and present more numerical evidences.

\subsection{Derivations of Eq. (4)}

Consider a perturbed nanoresonator driven by an incident electric field $\Ev_{\rm in} (\rv;\omega)$, and the scattered field $\Ev_{\rm sca} (\rv;\omega)$ is induced. Using the scattered field formalism, $\Ev_{\rm sca}$ in the perturbed system satisfies
\begin{align}
\left[\nablav\times\mu_0^{-1}\nablav\times-\omega^2\bm\varepsilon_{\rm p}(\rv;\omega)\right]\Ev_{\rm sca}(\rv;\omega)=\omega^2
\left[\bm\varepsilon_{\rm p}(\rv;\omega)-\bm\varepsilon_{\rm bg}(\omega)\right]\Ev_{\rm in},\nonumber
\end{align}
To manifest the permittivity difference between the perturbed and unperturbed systems, the above equation can be reformulated as
\begin{align}
\left[\nablav\times\mu_0^{-1}\nablav\times-\omega^2\bm\varepsilon_{\rm u}(\rv;\omega)\right]\Ev_{\rm sca}(\rv;\omega)=\omega^2
g(\rv)\Delta\bm\varepsilon  (\omega)   \Ev_{\rm in}+
\omega^2 f(\rv)
\Delta\bm\varepsilon (\omega)  \Ev_{\rm tot},
\label{eq:Esca_For}
\end{align}
with filling function $f(\rv)$ defined in Sec. \ref{Sec:LSFormalism}, another filling function $g(\rv)$ having a value of $1$ in the volume of the unperturbed nanoresonator, $V_{\rm res}$, and a value of $0$ elsewhere, and total field $\Ev_{\rm tot}=\Ev_{\rm in}+\Ev_{\rm sca}$. Equation \eqref{eq:Esca_For}, together with the Green's tensor of the unperturbed system defined in Eq. \eqref{eq:E-green}, leads to a self-consistent integral equation for $\Ev_{\rm sca}$,
\begin{align}
\Ev_{\rm sca}(\rv)=\omega^2\int \greenOPF_{\rm u}(\rv,\rv';\omega) f(\rv')\Delta\bm\varepsilon (\omega)\Ev_{\rm tot}(\rv') d^3\rv'+
                   \omega^2\int \greenOPF_{\rm u}(\rv,\rv';\omega) g(\rv')\Delta\bm\varepsilon (\omega)\Ev_{\rm in}(\rv') d^3\rv'.
                   \label{eq:Esca_For2}
\end{align}
Then, we transform the first (volume-integral) term on the right hand side of Eq. \eqref{eq:Esca_For2} to a surface integral over $\partial V_{\rm res}$ using the extrapolation technique detailed in Sec. \ref{Sec:DerEq1}, and obtain that
\begin{align}
\Ev_{\rm sca}(\rv)=    \omega^2 \oint_{\partial V_{\rm res}}  & \greenOPF_{\rm u}(\rv,\rvb'-\delta[h];\omega) \mathbf P_{\rm Geom}(\rvb') d^2\rvb'+\nonumber\\
                    +  \omega^2 \int_{V_{\rm res}}  & \greenOPF_{\rm u}(\rv,\rv'; \omega) \Delta\bm\varepsilon(\omega)\Ev_{\rm in}(\rv') d^3\rv',
\label{eq:Esca_For3}
\end{align}
where $\mathbf P_{\rm Geom}$ has the same expression as $\widetilde{\mathbf P}_{\rm Geom}$ in Eq. (1b) (main text) except that electric fields of perturbed modes, $\widetilde\Ev_{\rm p}$, need to be replaced by the total field $\Ev_{\rm tot}$.

We solve Eq. \eqref{eq:Esca_For3} by expanding $\Ev_{\rm sca}$ with unperturbed modes
\begin{align}
\Ev_{\rm sca}=\sum_{m}\beta_m\widetilde\Ev_{{\rm u}; n},
\end{align}
where $\beta_m$ denotes modal excitation coefficient to be determined. Following the similar routines as detailed in Sec. \ref{Sec:QNMEXP}, we derive that
\begin{subequations}
\begin{align}
{\bf\mathcal{H}}_0 \ket{\bm\beta}=\omega\left[\Iv+{\bf\mathcal{H}}_{\rm p}\right] \ket{\bm\beta}+\omega \left[\ket{\bm B}+ \ket{\bm S}\right],
\end{align}
with
\begin{align}
{\bf \mathcal {H}}_{{\rm p};n m}\simeq \mel{\left(\widetilde\Ev_{{\rm u};n}^{\rm bg}\right)^*}{
h \bm\Delta\varepsilon (\widetilde\omega_{{\rm u}; m})
}{\widetilde\Ev_{{\rm u};m}^{\rm res}}_{\scriptscriptstyle \partial V_{\rm res}},\\
S_n\simeq \mel{\left(\widetilde\Ev_{{\rm u};n}^{\rm bg}\right)^*}{
h \bm\Delta\varepsilon (\widetilde\omega_{{\rm u}; n})
}{\Ev_{\rm in}}_{\scriptscriptstyle \partial V_{\rm res}},\\
B_n =  \int_{V_{\rm res}} \widetilde\Ev_{{\rm u};n}(\rv)  \Delta\bm\varepsilon(\widetilde\omega_{{\rm u}; n})  \Ev_{\rm in}(\rv) d^3\rv,
\end{align}
\label{eq:beta}
\end{subequations}
where $\ket{\bm\beta}\equiv [\beta_1;\beta_2;\cdots]$, $\ket{\bm B}\equiv [B_1;B_2;\cdots]$ and $\ket{\bm S}\equiv [S_1;S_2;\cdots]$.

\subsection{Numerical Results}

\subsubsection{Influence of zero-frequency modes on numerical accuracy}

From many numerical tests of optical responses bwith Eq. (4), we observe that, even though zero-frequency modes have negligible impact on resonance positions and resonance linewidth of spectral responses---that is, they are irrelevant with spectral line shapes---, they contribute noticeably to magnitudes of observables. This relates to that the first-order correction to an eigenstate of a perturbed mode, in principle, requires counting the contribution from all
other modes including zero-frequency modes. While for predicting perturbed modal frequencies that determine spectral line shapes, zero-frequency modes, which are sufficiently gapped with the dominantly excited QNMs that
cover the frequency range of interest, are less important.

\begin{figure}[!htp]
\centering
\includegraphics[width=12cm]{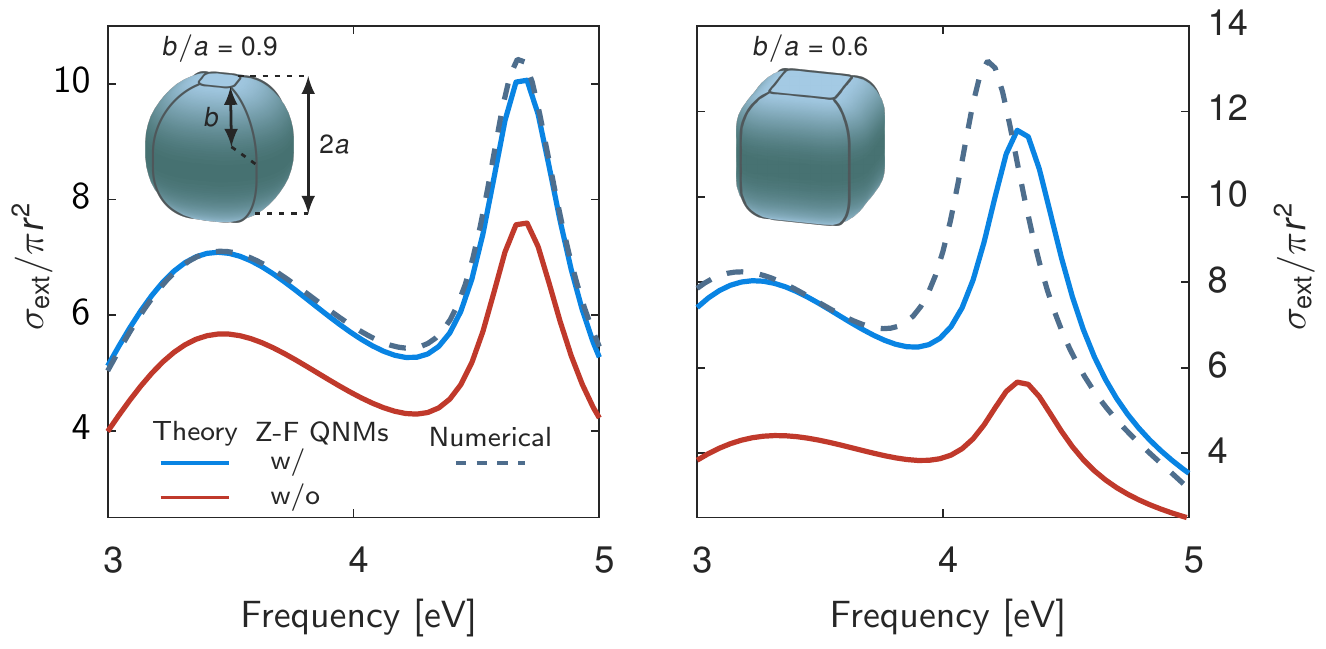}
\caption{{\bf Impact of zero-frequency (Z-F) modes on numerical accuracy of extinction-cress-section spectra reconstruction with Eq. (4)}. A silver nanosphere in air is deformed into cuboids, see Fig. 3 in the main text for parameter details. The theoretical results include 30 transverse QNMs, and Z-F modes are included. The numerical results are obtained with the boundary element method.
}
\label{Fig:ZF}
\end{figure}

As an example to illustrate the role of zero-frequency modes in reconstructing optical responses, we revisit the numerical case studied in Fig. 3(c) in the main text: A silver nanosphere in air is deformed into cuboids (parameterized by aspect ratio $b/a$). We compute extinction-cross-section spectra of two different cuboids---a small deformation with $b/a=0.9$ and a large deformation with $b/a=0.6$---, with Eq. (4) retaining 30 transverse QNMs. The computed theoretical results are compared with the numerical results in Fig. \ref{Fig:ZF}. We see that, independent of whether zero-frequency modes are included or not, the spectral shapes predicted with Eq. (4) agree reasonably with the numerical results particularly for the small deformation case. Therefore, for interpreting the physics due to modal interferences, zero-frequency modes are unimportant. However, the theoretical results of the extinction-cross-section spectra reconstructed without zero-frequency modes exhibit spectrally flat offsets with respect to the numerical results. In order to fix such offsets, we need to include zero-frequency modes. For all the numerical results demonstrated below, the contributions from
zero-frequency modes are included using the technique introduced in Sec. \ref{Sec:TRemarksLM}.

\subsubsection{Examples}

We now further evidence the validity of Eq. (4) with more numerical results. We consider a nanosphere in air and assume that the sphere is deformed into spheroids and cuboids. We consider that the nanosphere is made of two types of materials: (1) silver with a Drude permittivity $\varepsilon_{\rm metal}=1-\omega_{\rm p}^2/(\omega^2+i\omega\gamma)$ with $\hbar\omega_{\rm p}=9$ and $\gamma=0.0023\omega_{\rm p}$; (2) silicon with $\varepsilon_{\rm si}=12.96$.
The Drude nanosphere has a diameter of 100 nm; the Si nanosphere has a diameter of 400 nm. The transverse modes used in for reconstructing optical responses for the silver and silicon nanoresonators are plotted in Fig. 3(a) (main text) and Fig. S3(a), respectively.

\begin{figure}[!htp]
\centering
\includegraphics[width=12cm]{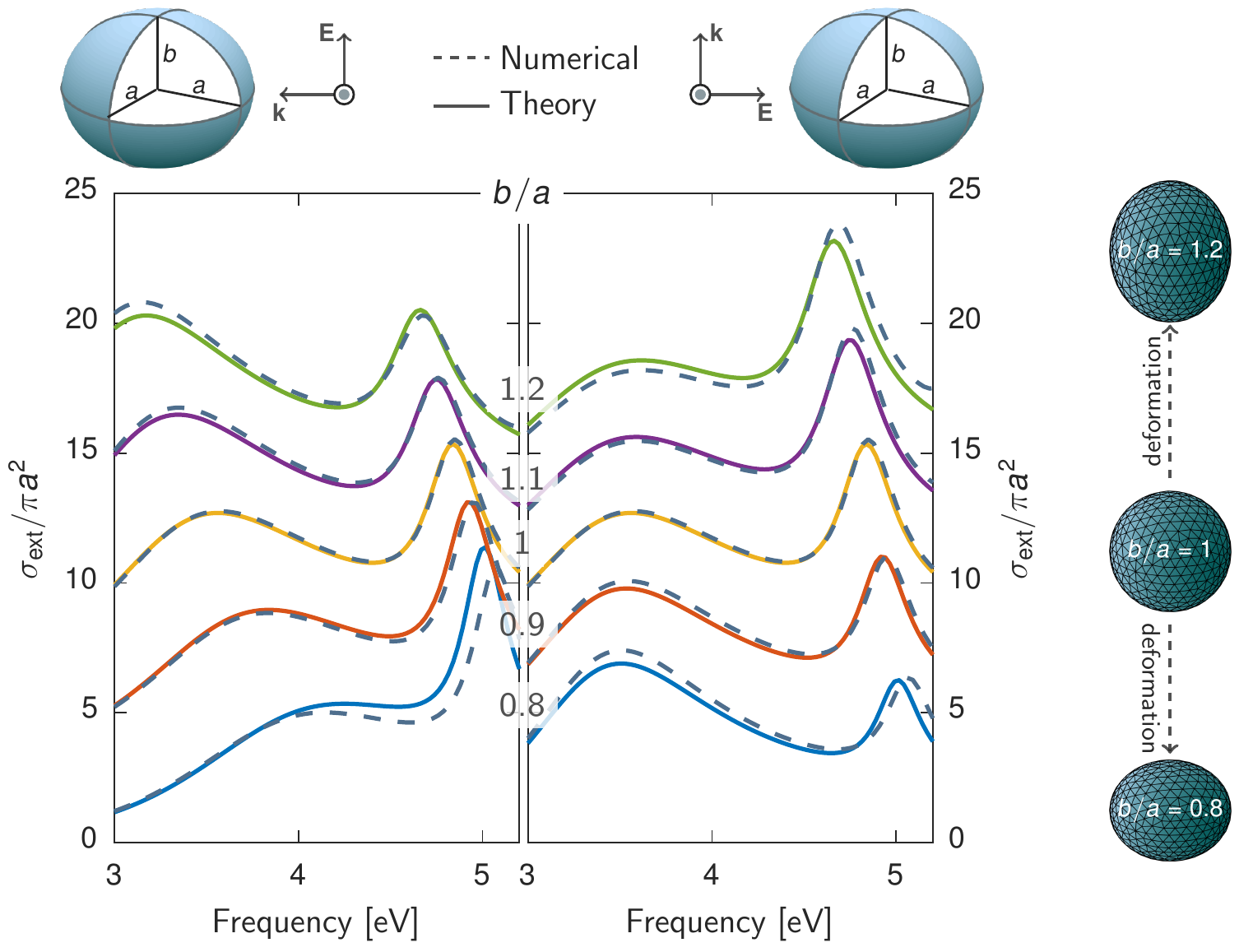}
\caption{{\bf Extinction-cross-section spectra of silver spheroids reconstructed with the perturbation theory}. The spheroids have a Drude permittivity $\varepsilon_{\rm metal}=1-\omega_{\rm p}^2/(\omega^2+i\omega\gamma)$ with $\hbar\omega_{\rm p}=9$ and $\gamma=0.0023\omega_{\rm p}$. The spheroids with selected aspect ratios $b/a=0.8,0.9,1.1,1.2$ ($a=50\,\rm nm$) are deformed from a nanosphere with a diameter of 100 nm.
}
\label{Fig:SiSpExt_Ag}
\end{figure}

\begin{figure}[!htp]
\centering
\includegraphics[width=12cm]{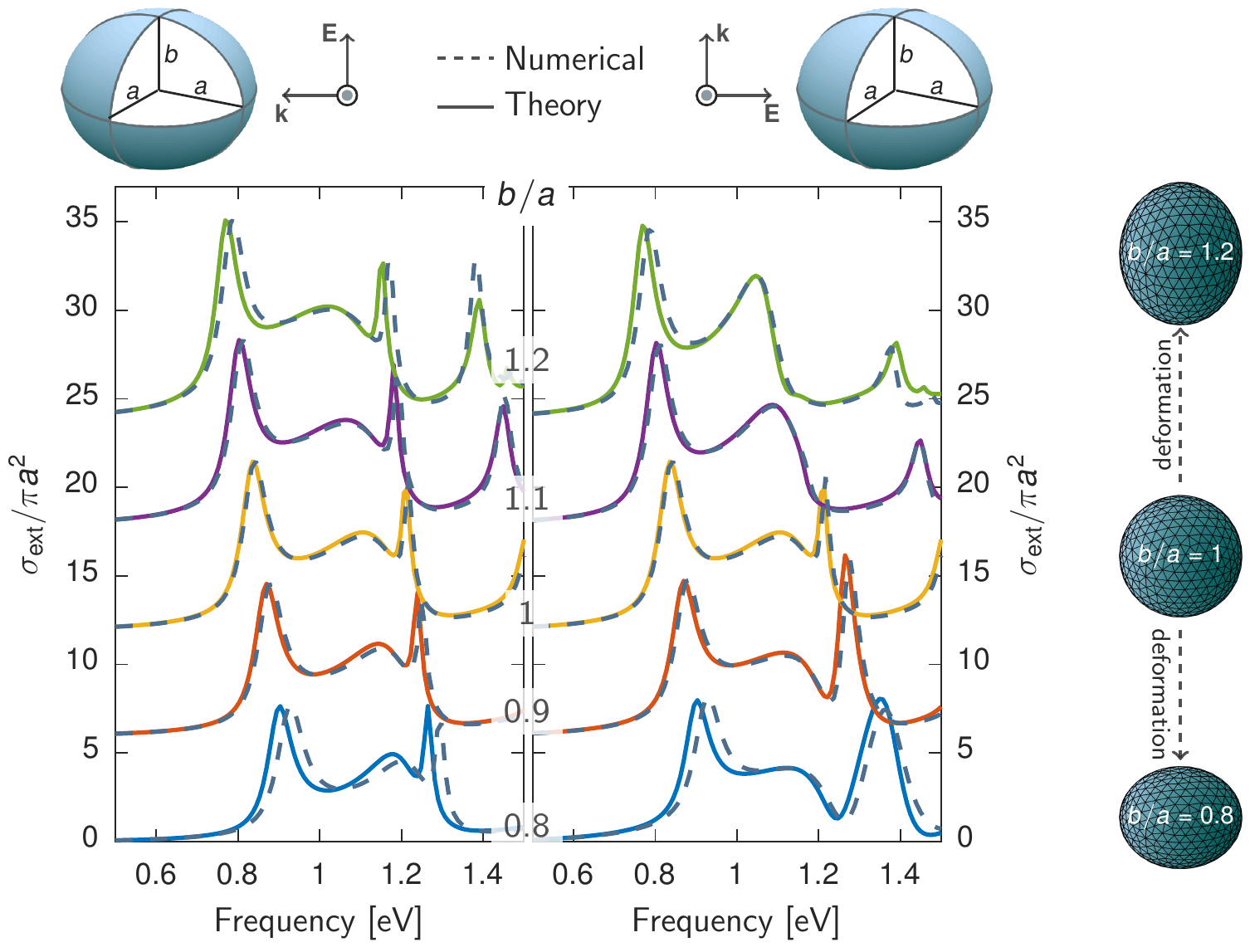}
\caption{{\bf Extinction-cross-section spectra of silicon spheroids reconstructed with the perturbation theory}. The spheroids have a permittivity $\varepsilon_{\rm si}=12.96$. The spheroids with selected aspect ratios $b/a=0.8,0.9,1.1,1.2$ ($a=200\,\rm nm$) are deformed from a nanosphere with a diameter of 400 nm.
}
\label{Fig:SiSpExt_Si}
\end{figure}

%\begin{figure}[!htp]
%\centering
%\includegraphics[width=8.5cm]{Fig_Cube_Ag_Extinction_Tikz.pdf}
%\caption{{\bf Extinction-cross-section spectra of metallic cubes reconstructed with the QNM perturbation theory}. The cubes have a Drude permittivity $\varepsilon_{\rm metal}=1-\omega_{\rm p}^2/(\omega^2+i\omega\gamma)$ with $\hbar\omega_{\rm p}=9$ and $\gamma=0.0023\omega_{\rm p}$. The cubes with selected aspect ratios $b/a=0.5,0.6,\cdots,0.9$ ($2a=100\,\rm[nm]$) are deformed from a nanosphere with a diameter of 100 [nm]. The theory results (sold curves) and the numerical results (dashed curves) exhibit an overall quantitative agreement.
%}
%\label{Fig:SiCbExt_Ag}
%\end{figure}

\begin{figure}[!htp]
\centering
\includegraphics[width=10cm]{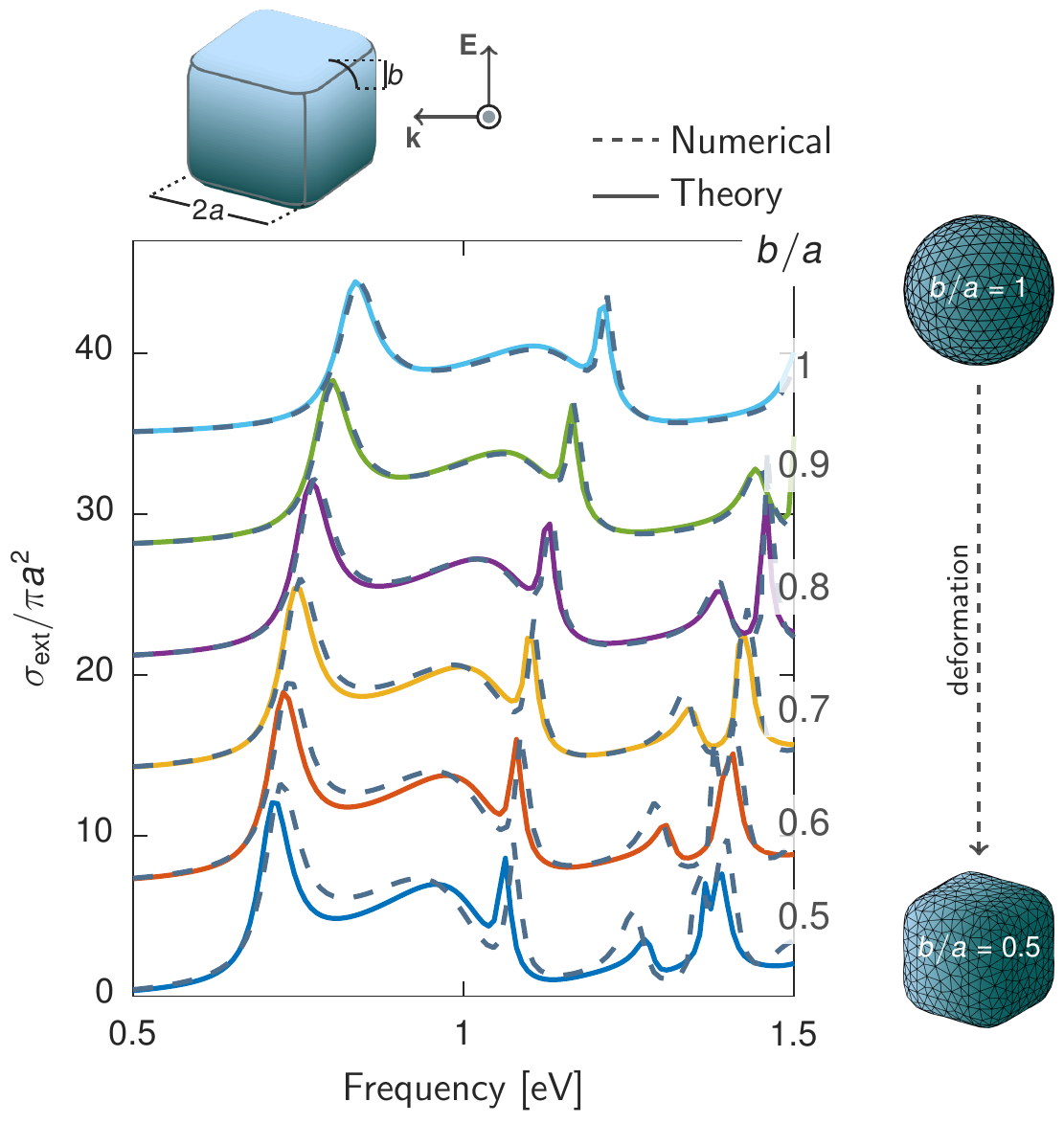}
\caption{{\bf Extinction-cross-section spectra of silicon cubes reconstructed with the perturbation theory}. The cubes have a permittivity $\varepsilon_{\rm si}=12.96$. The cubes with selected aspect ratios $b/a=0.5,0.6,\cdots,0.9$ ($a=200\,\rm nm $) are deformed from a nanosphere with a diameter of 400 nm.
}
\label{Fig:SiCbExt_Si}
\end{figure}

{\bf Spheres deformed into spheroids---}Figures \ref{Fig:SiSpExt_Ag} and \ref{Fig:SiSpExt_Si} plot the extinction-cross-section spectra of the silver and silicon spheroids, respectively. The spheroids are illuminated by plane waves that are polarized along one of two equal principal axes (right panels) and along the only different principal axes (left panels). We observe that the extinction-cross-section spectra computed with the perturbation theory agree well with the numerical results (obtained with the boundary-element-method) in terms of positions, widths, and heights of resonance peaks.

{\bf Spheres deformed into cuboids---}Figure~\ref{Fig:SiCbExt_Si} plots the extinction-cross-section spectra of the silicon cuboids, respectively. Due to high symmetry of cubes, we only consider cubes illuminated by plane waves that are polarized one of three directions of their sides. We again observe that the extinction-cross-section spectra computed with the perturbation theory agree well with the numerical results.
\begin{figure}[!h]
\centering
\includegraphics[width=12cm]{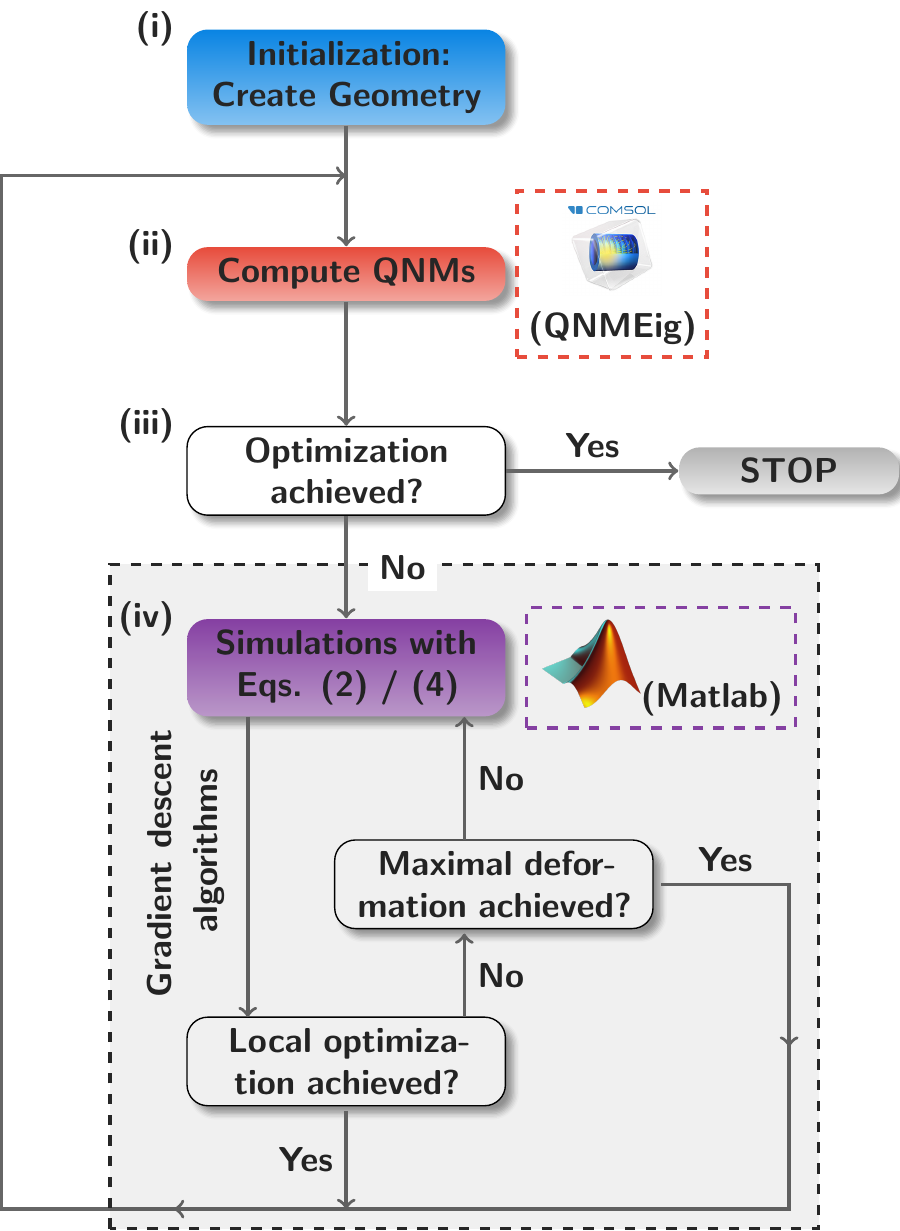}
\caption{{\bf Schematic implementation of inverse design using the QNM PT}. The procedures are summarized as follows.
(i) Start with a guessed nanoresonator. (ii) Compute QNMs of the nanoresonator covering the frequency range of interest, e.g., with the QNMEig solver~\cite{Yan:2018}. (iii) Determine whether the functionalities of the current nanoresonator meet the defined optimization goal. If yes, then stop otherwise continue to the next step. Optical responses of the nanoresonator---that are needed to be computed---are reconstructed with the QNM-expansion technique~\cite{Lalanne:2018}
(iv) Compute QNM frequencies and optical responses of deformed nanoresonators with Eqs. (2) or (4) using a few QNMs of the initial nanoresonator, and identify a local-optimized nanoresonator by searching the minimal mismatching between estimated functionalities of deformed nanoresonators and targeted ones. This step includes two ingredients: a loop for searching the local optimization, e.g., with the gradient descent approach; a checkpoint for judging whether the updated deformations reach the user-defined maximal values.
Finally, update the nanoresonator to the optimized one and start the next iteration.
}
\label{Fig:WorkFlow}
\end{figure}

\section{Workflow of Inverse Design}

One important application of the developed PT is in inverse design of nanoresonators. The inverse design is generally computationally expensive when parameter space has a large dimension and multi-optical-functionalities are targeted simultaneously. It, thus, calls
for improvement of numerical efficiency of full-wave simulations, and therein lies in the advantage of the PT---outstanding numerical efficiency empowered by that multiple full-wave computations over repeated frequency and wave-excitation variations are replaced by a single computation of a few dominant QNMs and, then by solving small-dimensional matrix problems of Eqs. (2) or (4) (main text). A general workflow for inverse design using the PT is outlined below and sketched in Fig. \ref{Fig:WorkFlow}.

%Starting with a gussed nanoresonator, we compute its QNMs covering the frequency range of interest. Then, we deform the nanoresonator (parameterized by a few user-defined parameters), whose QNM frequencies and optical responses under external stimuli are efficiently estimated with  Eqs. (2) or (4) using a few QNMs of the initial nanoresonator. An optimized nanoresonator is identified by searching the minimal mismatching---between estimated functionalities of deformed nanoresonators and targeted ones---in parameter space, by exploiting nonlinear optimization algorithms, such as gradient descent algorithms. The iterations are terminated until the optimized nanoresonator meets the targeted functionalities.

\bibliographystyle{apsrev4-1}
\bibliography{supplement}